\documentclass[prd,aps,preprint,amsmath,nofootinbib,amssymb,eqsecnum,showkeys,tightenlines]{revtex4-1}
\usepackage{slashed}
\usepackage{epsfig,latexsym,cancel,amssymb,amsmath,verbatim,mathrsfs}
\usepackage{color}
\usepackage{graphicx}

\def\ra{\rightarrow}
\def\L{\left(}
\def\R{\right)}
\def\wt{\widetilde}
\def\Ld{\Lambda}
\def\ld{\lambda}
\def\f{\frac}
\newcommand{\be}{\begin{equation}}
\newcommand{\ee}{\end{equation}}
\newcommand{\bea}{\begin{eqnarray}}
\newcommand{\eea}{\end{eqnarray}}
\newcommand{\ba}{\begin{array}}
\newcommand{\ea}{\end{array}}

\long\def\symbolfootnote[#1]#2{\begingroup%
\def\thefootnote{\fnsymbol{footnote}}\footnote[#1]{#2}\endgroup}

\newcommand{\beq}{\begin{equation}}
\newcommand{\eeq}{\end{equation}}

\begin{document}

\title{$(g-2)_\mu$ Versus Flavor Changing Neutral Current Induced by\\  the Light $(B-L)_{\mu\tau}$ Boson}

\author{Zhaofeng Kang}
\email[E-mail: ]{zhaofengkang@gmail.com}
\affiliation{School of Physics, Huazhong University of Science and Technology, Wuhan 430074, China}



\author{Yoshihiro Shigekami}
\email[E-mail: ]{sigekami@post.kek.jp}
\affiliation{School of Physics, Huazhong University of Science and Technology, Wuhan 430074, China}

\date{\today}

\begin{abstract}

We propose the local $(B-L)_{\mu\tau}$ model, which minimally retains the local $B-L$ extension for the sake of neutrino phenomenologies, and at the same time presents an invisible gauge boson $Z'$ with mass $\sim {\cal O}(10)$ MeV to account for the discrepancy of the muon anomalous magnetic moment. However such a scenario is challenged by flavor physics. To accommodate the correct pattern of Cabibbo-Kobayashi-Maskawa matrix, we have to introduce either a $SU(2)_L$ doublet flavon or vector-like quarks plus a singlet flavon. In either case $Z'$ induces flavor changing neutral current (FCNC) in the quark sector at tree-level. We find that the former scheme cannot naturally suppress the FCNC from the down-type quark sector and thus requires a large fine-tuning to avoid the stringent $K \to \pi \nu\bar \nu$ bound. Whereas the latter scheme, in which FCNC merely arises in the up-type quark sector, is still free of strong constraint. In particular, it opens a new window to test our scenario by searching for flavor-changing top quark decay mode $t\ra u/c+$(invisible), and the typical branching ratio $\sim\mathcal{O}(10^{-4})$.

\end{abstract}


\maketitle

\section{Introduction}

The non-vanishing neutrino masses is a clean signature of physics beyond the particle standard model (SM). The gauge group extension by $U(1)_{B-L}$ \cite{Davidson:1978pm,Mohapatra:1980qe,Marshak:1979fm} provides an elegant way to understand the origin of neutrino masses: Right handed neutrinos (RHNs) are indispensable to cancel anomaly associated with $U(1)_{B-L}$, and they are supposed to gain Majorana masses after spontaneously breaking of  $U(1)_{B-L}$, thus fulfilling the  the seesaw mechanism~\cite{Seesaw1,Seesaw2,Seesaw3,Seesaw4,Seesaw5,Seesaw6,Seesaw7}. Conventionally, three RHNs are introduced under the assumption that  all  three families of SM fermions are charged universally under  $U(1)_{B-L}$. However, this assumption is not necessary neither from gauge anomaly cancellation principle nor from neutrino physics: That cancellation applies to each family, and two RHNs is sufficient to reproduce the correct neutrino mass and mixing patterns observed by the present experiments (see new results of Planck collaboration \cite{Aghanim:2018eyx} and the global fit \cite{Esteban:2018azc}).

As a matter of fact, relaxing that assumption may offers a big bonus, namely explaining the long-standing $\sim 3.7\sigma$  discrepancy between the experimental values and the SM predictions for the muon anomalous magnetic moment $(g-2)_{\mu}$~\cite{Hagiwara:2011af,Keshavarzi:2018mgv,Bennett:2006fi,Roberts:2010cj,Davier:2010nc,Davier:2017zfy,Davier:2019can}. If interpreted by light particles below the weak scale, the idea initiated in Ref.~\cite{darkA}, one needs a muon-philic but electron-phobic boson. Otherwise, one has to very carefully set up the light world so as to avoid a bunch of strong exclusions, which are available from the electron-related  low energy experiments; see a recent analysis on the dark photon case~\cite{darkA:new}. Such a status drives us to consider the hypothesis that only the second and third families of SM fermions are charged under $U(1)_{B-L}$,\footnote{In order to  explain anomalous rare $B$ decays,  $U(1)_{(B-L)_3}$, under which only the third family of SM fermions are charged, was proposed by Ref.~\cite{Alonso:2017uky}, focusing on the TeV scale gauge boson. Such kind of extension was first proposed by Ref.~\cite{Babu:2017olk}, and recently, the LHC constraints on the $U(1)_{(B-L)_3}$ gauge boson are discussed by Ref.~\cite{Elahi:2019drj}.} and the resulting gauge group is denoted as $(B-L)_{\mu\tau}$. Set the quarks aside,  the gauge boson $Z'$ in our model assembles the one in the local $L_\mu-L_\tau$ model~\cite{LmuLtau1,LmuLtau2} and thus is able to explain the $(g-2)_{\mu}$ as in the latter~\cite{LmuLtau:g-2}. Although there are some constraints from the CCFR \cite{CCFR}, BaBar \cite{BABAR} and Big Bang Nucleosynthesis (BBN) \cite{BBN1,BBN2,BBN3} in the light mass region,\footnote{Other constraints in various anomaly free $U(1)$ model can be found in Ref. \cite{Bauer:2018onh}, including future experimental prospects.} the $Z'$ mass about $\mathcal{O}(10)$-$\mathcal{O}(100)$ MeV still have favored regions to explain its deviation within $2\sigma$. There are many discussions about phenomenology of this mass region in the $U(1)$ extended model, see for example Refs. \cite{Davoudiasl:2014kua,Fuyuto:2014cya,Jeong:2015bbi,Fuyuto:2015gmk,Kaneta:2016vkq,Datta:2017pfz,Araki:2017wyg,Gninenko:2018tlp,Kamada:2018zxi,Biswas:2019twf}. Note that there are also similar discussions for the case of a little bit heavy $Z'$ mass ($> 1 \mathchar`- 10$ GeV), see for example \cite{Altmannshofer:2016jzy,Ko:2017quv,DiChiara:2017cjq,Bonilla:2017lsq,Bian:2017rpg,Arcadi:2018tly,Falkowski:2018dsl,Chun:2018ibr,Hutauruk:2019crc}.

Concerning the quark sector, there is a problem of accommodating the correct pattern of the Cabibbo-Kobayashi-Maskawa (CKM) matrix in our setup since there are no Yukawa couplings between the first and the other generations. One of the procedures to resolve this is to introduce a flavon field ${\cal F}$, and two schemes are available depending on the representations of the flavon: 
\begin{itemize}
\item  If ${\cal F}$ is a $SU(2)_L$ Higgs doublet,  all quark Yukawa couplings can be write down at the renormalizable level, recovering the SM Yukawa structure after ${\cal F}$ developing a VEV. 
\item If ${\cal F}$ is a SM singlet scalar field, at least one vector-like quarks are also needed. In this case, we obtain all elements of Yukawa coupling by integrating out these TeV scale vector-like quarks~\footnote{Introducing vector-like fermions has a possibility to obtain the hierarchies of the SM fermion masses in $SU(2)_L \times SU(2)_R \times U(1)_{B-L}$ models \cite{Berezhiani:1983hm,Chang:1986bp,Rajpoot:1987fca,Rajpoot:1986nv,Rajpoot:1987ji,Davidson:1987mh,Rajpoot:1988gx,Berezhiani:1991ds}.}.
\end{itemize}
In either scheme, due to the off-diagonal elements of Yukawa couplings and flavor dependent $U(1)_{B-L}$ charges of quarks, $Z'$ couplings of quarks become flavor violating ones. Therefore, we should take into account the constraints from the CKM matrix.

The two schemes may give rise to significantly different patterns of flavor changing neutral current (FCNC). In the second scheme, by choosing quantum numbers of these vector-like quarks, the mixings in the down-type quark sector are present only within the second and third families, and as a consequence FCNC mediated by $Z'$ is absent in the down-type quark sector. It helps to evade the extremely severe constraints from the $B \to K Z'(\ra \nu \bar{\nu})$ and $K \to \pi Z'(\ra \nu \bar{\nu})$ processes. By contrast, they are present in the first scheme which hence is clearly excluded. We point out that in the first scheme one can establish a connection between $(g-2)_\mu$ and flavor-changing top quark decay, $t \to q Z'$ where $q = u, c$.

This paper is organized as follows: In the Section \ref{sec:B-Lmodel} we introduce our flavored $U(1)_{B-L}$ model. Then we discuss some phenomenologies induced by our $Z'$ in the Section \ref{sec:ZpFCNC}. In this section, we mainly discuss $(g-2)_{\mu}$, flavor-changing top quark decay and $K$ and $B$ meson decays. The Section \ref{sec:summary} gives the summary of our paper.

\section{Gauged $({B-L})_{\mu\tau}$ with a light gauge boson}
\label{sec:B-Lmodel}

In this section we present the ingredients for building a realistic local $({B-L})_{\mu\tau}$ model that is capable of providing a light gauge boson at the sub-GeV scale and moreover producing the observed fermion masses and mixings. 

\begin{table}[]
\begin{center}
\begin{tabular}{|c|c|c|c|c|}
\hline
~ ~~ ~~ ~~ ~&~~\rm spin~ ~& ~~$SU(2)_L$ ~~& ~~$U(1)_Y$~~ & ~~ $({B-L})_{\mu\tau}$~~  \\
\hline
$Q_i$ & $1/2$ &     2  &    1/6  &    1/3  \\
$u_{R,i}$ & $1/2$ &     1  &    2/3  &    1/3  \\
$d_{R,i}$ & $1/2$ &     1  &   $-1/3$  &    1/3  \\
\hline\hline
$U_L$ & $1/2$ &    1 &    2/3  &   1/3 \\
$U_R$ & $1/2$ &    1 &    2/3  &   1/3 \\
${\cal F}$ & $0$ &    1 &    0  &   1/3 \\
\hline\hline
$L_i$ & $1/2$ &     2  &    $-1/2$  &   $-1$ \\
$e_{R,i}$ &$1/2$ &     1&    $-1$ &   $-1$  \\
$N_{R,i}$ &1/2 &     1  &    0  &    $-1$   \\
\hline\hline
$H$ &$0$ &   2  &    1/2 &   0   \\
$\Phi$ &0 &     1  &    0  &   $2$ \\
\hline
\end{tabular}
\end{center}
\caption{Field content and quantum numbers in the minimal $({B-L})_{\mu\tau}$ extension; here $i=2,3$ and the first generation of fermions which are neutral under this new gauge group are not listed. In the doublet flavon model, vectro-like quarks are replaced by $H_{\mu\tau}$ which has the same quantum numbers as $H$ except for carrying 1/3 charge under  $({B-L})_{\mu\tau}$. \label{QN}}
\end{table}
\subsection{Model blocks}

In the gauged  $({B-L})_{\mu\tau}$ model, in order to produce the required phenomenologies, minimally we need new fields listed below:
\begin{description}
\item[ Right-handed neutrinos $N_R$]  They have  $({B-L})_{\mu\tau}$ charge $-1$, required to cancel anomalies associated with $({B-L})_{\mu\tau}$. Meantime they lead to active neutrino masses via the seesaw mechanism. To obtain realistic neutrino masses which allows one massless neutrino, at least two families of RHNs are needed. This fact explains why we consider the flavored $(B-L)$ model with two families of fermions charged under it. However, owing to the selection rule by $({B-L})_{\mu\tau}$, the correct pattern of neutrino mixings  requires additional extension, and we will comment on this in the following.

\item[$({B-L})_{\mu\tau}$ Higgs $\Phi$] To spontaneously break $({B-L})_{\mu\tau}$ and simultaneously give Majorana mass to RHNs carrying $({B-L})_{\mu\tau}$ charge $-1$, we need a corresponding Higgs field $\Phi$ carrying $({B-L})_{\mu\tau}$ charge $+2$. The relevant terms are
\begin{eqnarray}\label{seesaw}
-{\cal L}\supset Y^N_{ij} \overline{L}_i \wt HN_{R,j}+ \f{\ld^N_{ij}}{2}\Phi \overline{ N_{R,i}^C} N_{R,j}+ V(\Phi),
\end{eqnarray}
where the $({B-L})_{\mu\tau}$ Higgs sector $V(\Phi)$ is responsible for breaking $({B-L})_{\mu\tau}$ at a scale $\langle \Phi\rangle \equiv v_\phi / \sqrt{2}$.  However, Eq.~(\ref{seesaw}) is not able to produce the observed neutrino mixings: Gauge symmetry does not allow terms $\bar L_1 \wt H N_{R,i}$ with $i=2,3$, and hence the first generation of active neutrino and the others do not mixing. Although it is important to accommodate observed neutrino mixings, we treat this issue as a future work since it does not affect the quark FCNCs mediated by $Z'$.

\item[Flavon]  The $({B-L})_{\mu\tau}$ gauge symmetry forbids couplings between the first and the other generations fermions, and consequently the minimal model fails in accommodating the CKM and Pontecorvo-Maki-Nakawaga-Sakata (PMNS) matrices in the quark and neutrino sectors, respectively. In order to  generate the correct CKM/PMNS matrices, flavons carrying proper quantum numbers are indispensable. Let us focus on the quark sector in this paper, and the discussions can be easily generalized to the neutrino sector. The concrete flavon depends on how we realize it. For instance, we may choose the flavon ${\cal F}_{1/3}$, a $SU(2)_L$ singlet having  $({B-L})_{\mu\tau}$  charge $1/3$; the subscript will be dropped. Then the quark Yukawa sector at $d=5$ level reads
\begin{eqnarray}\label{mixing:Q}
-{\cal L}_q\supset  Y_{ij}^d \overline{Q}_i H d_{R,j}+ Y_{11}^d \overline{Q}_1 H d_{R,1}+ Y_{ij}^u \overline{Q}_i\wt H  u_{R,j}+ Y_{11}^u \overline{Q}_1\wt H  u_{R,1}
\cr
+\f{y_{i1}^u}{\Ld} \overline{Q}_i\wt H  u_{R,1}{\cal F}+\f{y_{1i}^u}{\Ld} \overline{Q}_1\wt H  u_{R,i}{\cal F}^*+h.c.,
\end{eqnarray}
It is seen that after the flavon developing a VEV $\langle {\cal F}_{1/3}\rangle \equiv v_f / \sqrt{2}$, the $d=5$ operators just give rise to the desired quark mixing.

In the above Yukawa sector we have set $y_{i1}^d\ra 0$ to avoid large quark mixings in the down-type quark sector which has been stringently constrained. More details will be given later. The $ad~ hoc$ structure  $y_{i1}^d\ra 0$ is naturally understood in the renormalizable realization of the $d=5$ operators, if we merely introduce a pair of heavy vector-like quarks $(U_L, U_R)$, which transform identically with the $SU(2)_L$ singlet quarks $u_{R,i}$ and thus have the most generic renormalizable Lagrangian 
\begin{eqnarray}\label{UV}
-{\cal L}_{U}=  M_U \overline{U}_L U_R+M_{Ui} \overline{U}_L u_{R,i}+\ld_1 \overline{U}_L u_{R,1}{\cal F}+\ld_{i} \overline{Q}_i\wt H  U_R+h.c.
\end{eqnarray}
Replacing ${\cal F}$ with its VEV and defining $M_{U1}=\ld_1 v_f / \sqrt{2}$, we obtain the effective theory for the light quarks after integrating out the heavy quarks via the equation of motion,
\begin{eqnarray}\label{}
U\approx \f{1}{M_U}M_{Ua}P_Ru_a+\f{M_{Ua}}{M_U^2}i\gamma_\mu\partial^\mu P_R u_a+ \f{1}{M_U}\ld_i^*P_L\wt H^\dagger Q_i,
\end{eqnarray}
with $a=1,2,3$ while $i=2, 3$. Substituting it into the terms involving light-heavy couplings in the Lagrangian Eq.~(\ref{UV}), one gets 
\begin{eqnarray}\label{eff}
{\cal L}_{eff}\supset c_{ab}\bar u_a i\gamma_\mu\partial^\mu P_R u_b+ Y_{ib} \overline{Q}_i\wt HP_Ru_b+h.c.
\end{eqnarray}
where the coefficients/couplings are given by  
\begin{eqnarray}\label{Ecouplings}
c_{ab}=\f{M_{Ua}M_{Ub}^*}{M_U^2},\quad Y_{ib}=\ld_i\f{M_{Ub}}{M_U}.
\end{eqnarray}
So, the resulting effective Lagrangian includes the kinetic mixing for the three generations of right-handed up quarks and mass mixings (after the electroweak symmetry breaking (EWSB)),  in particular $\ld_1 M_{U1}/M_U v\bar u_{L,1}u_{b,i}$. Before diagonalizing the quark mass matrix, one should choose a basis in which the kinetic terms are canonical. This is done through the unitary rotation $u_{R}\ra \wt W_u u_{R}$. Then, the terms $Y_{1b} \overline{Q}_1\wt H P_R u_b$ which is absent in the second term in Eq.~(\ref{eff}) is generated. Note that $M_{U1}\propto v_f\lesssim 100$ GeV from the later analysis below Eq.~(\ref{Zmass}), but the size of $M_{Ui}$ is not subject to this bound. Anyway, as long as $M_U$ is not far above $v_f\sim 100$ GeV, the  size of the couplings given in Eq.~(\ref{Ecouplings}) can be sufficiently large to accommodate CKM. On the other hand, the light colored vector-like quarks, says around the TeV scale, can be abundantly produced at the hadronic colliders and decay into the SM quarks, leaving signals at the LHC. But the concrete signatures depends on couplings and need a specific analysis elsewhere.

Alternatively, the flavon can be $SU(2)_L$ Higgs doublet $H_{\mu\tau}$ with $({B-L})_{\mu\tau}$ charge  1/3, which admits Yukawa couplings between the first and second/third families, 
\begin{eqnarray}\label{F:doublet}
-{\cal L}_Y\supset\wt {Y}^u_{1i}\overline{Q}_1 \wt H_{\mu\tau} u_{R,i}+
\wt{Y}^d_{i1} \overline{Q}_i H_{\mu\tau} d_{R,1}+h.c.,
\end{eqnarray}
where the terms $\overline{Q}_i \wt H_{\mu\tau} u_{R,1}$ and $\overline{Q}_1 H_{\mu\tau} d_{R,i}$ are  forbidden by $(B-L)_{\mu\tau}$. Here we do not need to introduce vector-like quarks. However, in general it is unnatural to suppress the $Z'$-mediated FCNC from the down-type quark sector.

\end{description}
We summarize the  field content in table~\ref{QN}.

\subsection{Bosonic particle mass spectra}

\subsubsection{Singlet flavon model}

From the above statement we know that in the minimal model the SM Higgs sector is extended by two SM singlet scalar fields, the $B-L$ Higgs field $\Phi$ and the flavon ${\cal F}$~\footnote{ It is of interest to notice that $\Phi$ may be unnecessary in the presence of ${\cal F}$. In that case the $B-L$ is  spontaneously broken by ${\cal F}$ and as a consequence the right-handed neutrinos  are forced to form Dirac neutrinos with the left-handed neutrinos. }.  Charge assignments leads to a trivial multi dimensional Higgs potential,
\begin{eqnarray}\label{}
V(H,\Phi, {\cal F})&=&\L-m_H^2|H|^2+\ld_H|H|^4\R+\L-m_{\Phi}^2|\Phi|^2+\ld_\phi|\Phi|^4\R+\L-m_{{\cal F}}^2|{\cal F}|^2+\ld_{\cal F}|{\cal F}|^4\R \nonumber \\
&&+\ld_{H\Phi}|H|^2|\Phi|^2+\ld_{H{\cal F}}|H|^2|{\cal F}|^2+\ld_{\Phi{\cal F}}|\Phi|^2|{\cal F}|^2. \label{eq:potential}
\end{eqnarray}
Let us write $S=\f{v_s+s+iA_s}{\sqrt{2}}$ ($S = H^0, \Phi, {\cal F}$), where $H^0$ is the neutral component of $H$. $v_{h}=246$ GeV generates the electroweak scale, while $v_{\phi}$ and $v_{f}$ will be discussed later. 

After breaking $SU(2)_L \times U(1)_Y \times U(1)_{(B-L)_{\mu\tau}} \to U(1)_{\rm em}$, four physical states are remaining in the scalar sector: three CP-even neutral Higgs bosons $h, \phi, f$ and one CP-odd neutral Higgs boson $A$. Note that the charged Higgs in $H$ and $A_h$ are eaten by the SM gauge bosons and $A_{\phi}$ is eaten by the new gauge boson, $Z'$. By applying conditions to give VEVs for all Higgs bosons, $\partial V / \partial S = 0$, we can obtain the masses for physical states. In particular, the squared mass matrix for CP-even Higgs bosons are given as
\begin{eqnarray}
V \supset \frac{1}{2} \begin{pmatrix} h & \phi & f \end{pmatrix} M^2 \begin{pmatrix} h \\ \phi \\ f \end{pmatrix}, \quad M^2 = \begin{pmatrix}
2 \ld_H v_h^2 & \ld_{H\Phi} v_h v_{\phi} & \ld_{H{\cal F}} v_h v_f \\
\ld_{H\Phi} v_h v_{\phi} & 2 \ld_{\Phi} v_{\phi}^2 & \ld_{\Phi{\cal F}} v_{\phi}  v_{f} \\
\ld_{H{\cal F}} v_h v_{f} & \ld_{\Phi{\cal F}} v_{\phi}  v_f & 2 \ld_{\cal F} v_{f}^2
\end{pmatrix}.
\end{eqnarray}
It is ready to obtain a SM-like Higgs boson by setting $\ld_{H\Phi}, \ld_{H{\cal F}}\ll 1$. They are irrelevant to the main line of this article, and thus we will not expand the discussions on them. 

In this model there is no mixing between the electroweak gauge bosons and $(B-L)_{\mu\tau}$ at tree level. The mass of the latter receives contributions both from the new Higgs field $\Phi$ and flavon ${\cal F}$:
\begin{align}\label{Zmass}
&|D_{\mu} {\cal F}|^2+ |D_{\mu} \Phi|^2 \supset \frac{1}{2}M_{Z'}^2Z_\mu'Z'^{\mu}, 
\end{align}
with the mass squared $M_{Z'}^2=g_{B-L}^2\L v_f^2/9+4v_\phi^2\R\sim {\cal O}(0.01{\rm GeV})^2- {\cal O}(0.1{\rm GeV})^2$ of interest in this paper.  Moreover, the required $g_{B-L}\sim 10^{-3}$ in this mass region to account for $(g-2)_\mu$; see Eq.~(\ref{10MeV}) and Eq.~(\ref{100MeV}). Hence, one has the rough upper bound on $v_f\lesssim 30{\rm GeV}-300{\rm GeV} $, which has some implications to the CKM as discussed before.

\subsubsection{Doublet flavon model}

If the flavon field is a $SU(2)_L$ Higgs doublet, $H_{\mu\tau}$ whose $B-L$ charge is 1/3, the situation becomes more complicated. First, the Higgs sector becomes a special version of two Higgs doublet model, forbidding the crucial $\mu$-term $\mu H H_{\mu\tau}$ to realize realistic gauge symmetries spontaneously breaking~\footnote{One cannot rely on the term $-m^2_{H_{\mu\tau}}|H_{\mu\tau}|^2$ to drive the non-vanishing VEV $\langle H_{\mu\tau}\rangle$, because it results in light physical Higgs bosons with mass $\sim \langle H_{\mu\tau}\rangle$ which have been excluded by experiments.}. To overcome this problem, we are forced to introduce an additional Higgs singlet which carries $B-L$ charge $+1/3$, the same as ${\cal F}$. It develops VEV and then generates the $\mu$-term via the trilinear term $ {\cal F}H_{\mu\tau}^{\dagger}H$.

In this model, because $H_{\mu\tau}$ contributes mass both to the weak gauge bosons and $B-L$ Higgs boson, $Z-Z'$ mixing arises at tree-level. The squared mass matrix for gauge bosons can be obtained from the covariant derivatives of $H$, $H_{\mu\tau}$ and $\Phi$ as
\begin{align}
&|D_{\mu} H|^2 + |D_{\mu} H_{\mu\tau}|^2 + |D_{\mu} \Phi|^2 \supset \frac{1}{2} \begin{pmatrix} B_{\mu} & W^3_{\mu} & {\hat{Z}'}_{\mu} \end{pmatrix} M_G^2 \begin{pmatrix} B^{\mu} \\ W^{3\, \mu} \\ \hat{Z}^{\prime \mu} \end{pmatrix}. 
\end{align}
As usual, $B_{\mu}$ and $g_Y$ are the gauge boson and its coupling of $U(1)_Y$, while $W^a_{\mu}$ and $g_2$ are those of $SU(2)_L$. Then the mass squared matrix is parameterized as the following
\begin{align}
& M_G^2 = \frac{v_h^2}{4} \begin{pmatrix}
g_Y^2 (1 + r_{\mu\tau}) & - g_Y g_2 (1 + r_{\mu\tau}) & \frac{2}{3} g_Y g_{B-L} r_{\mu\tau} \\
- g_Y g_2 (1 + r_{\mu\tau}) & g_2^2 (1 + r_{\mu\tau}) & - \frac{2}{3} g_2 g_{B-L} r_{\mu\tau} \\
\frac{2}{3} g_Y g_{B-L} r_{\mu\tau} & - \frac{2}{3} g_2 g_{B-L} r_{\mu\tau} & g_{B-L}^2 \left(\frac{4}{9} r_{\mu\tau} + 16 r_{\phi}\right) \label{eq:Zmassmatrix} \\
\end{pmatrix},
\end{align}
where $r_{\mu\tau} \equiv \frac{v_{\mu\tau}^2}{v_h^2}$ and $r_{\phi} \equiv \frac{v_{\phi}^2}{v_h^2}$; $v_{\mu\tau}(1+1/r_{\mu\tau})^{1/2}=246$ GeV with $v_{\mu\tau}$ the VEV of the second Higgs doublet, whose size is favored to be around the weak scale, giving rise to the desired CKM without large Yukawa couplings between $H_{\mu\tau}$ and quarks. By diagonalizing $M_G^2$, $\hat{Z}_{\mu}$ are given by
\begin{equation}
{\hat{Z}'}_{\mu} = u_{\hat{Z}A} A_{\mu} + u_{\hat{Z}Z} Z_{\mu} + u_{\hat{Z}Z'} Z'_{\mu},
\label{eq:Zmixing}
\end{equation}
with the elements of diagonalizing matrix of $M_G^2$, $u_{\hat{Z}A}$, $u_{\hat{Z}Z}$ and $u_{\hat{Z}Z'}$. Here, $A_{\mu}$ and $Z_{\mu}$ correspond to the SM photon and $Z$ boson. Since we are considering the light $Z'$ mass, $M_{Z'} \ll m_Z$ and moreover $g_{B-L}\ll 1$ to explain $(g-2)_{\mu}$, the mixing in Eq. \eqref{eq:Zmixing} are expected to be small. Additionally, we can set $r_{\mu\tau} \ll 1$ without loss of realization of desired $Z'$ mass. This induces an another suppression in the mixing, and therefore, we ignore flavor violating couplings due to $Z-Z'$ mixing.

In addition to the gauge sector, the Higgs sector has flavor violating couplings since not only CP-even Higgs but also charged Higgs are remaining as a physical mode. Although it is interesting to investigate the predictions and/or constraints on the model parameters from some flavor physics as like $B \to D^{(*)} \ell \nu$, the detailed study is beyond the scope of this paper. Actually, one can work in the parameter space region where the states in $H_{\mu\tau}$ are sufficiently heavy, suppressing the flavor changing from the Higgs sector.

\subsection{The CKM matrix}

Let us start from the more interesting singlet flavon model. After both $H$ and ${\cal F}$ developing VEVs, quarks gain Dirac mass terms $m_q^0 \bar q'_L q'_R$, and the resulting mass matrices take the form of
\begin{eqnarray}\label{mass:singlet}
m_u^0 = \f{v_h}{\sqrt{2}}
\begin{pmatrix}
Y^u_{11}  &  y^u_{12}v_f/\Ld  &  y^u_{13}v_f/\Ld \\
y^u_{21}v_f/\Ld   & Y^u_{22}  & Y^u_{23} \\
y^u_{31} v_f/\Ld &Y^u_{32} & Y^u_{33} \end{pmatrix}, ~~~ m_d^0 = \f{v_h}{\sqrt{2}}
\begin{pmatrix} Y^d_{11}  &  0  &  0 \\
0   & Y^d_{22}  & Y^d_{23} \\
0 &Y^d_{32} & Y^d_{33} \end{pmatrix}, \label{eq:quarkmass}
\end{eqnarray}
They are diagonalized through the usual bi-unitary transformations. They relate the quarks in the flavor (labelled with a prime) and mass basis as $q'^i_L= U_q^{ij}q^j_L$ and $q'^i_R= W_q^{ij}q^j_R$: $U_q^\dagger m_q^0(m_q^0)^\dagger U_q=m_q^2$ and $W_q^\dagger (m_q^0)^\dagger m_q^0 W_q=m_q^2$ with $m_q$ giving positive quark masses. The Cabibbo-Kobayashi-Maskawa (CKM) matrix is defined as
\begin{eqnarray}\label{}
V_{\rm CKM}=U_u^\dagger U_d= \left(\begin{array}{ccc} V_{ud} & V_{us} & V_{ub}
 \\
 V_{cd} & V_{cs} & V_{cb}
 \\
 V_{td} & V_{ts} &  V_{tb}\end{array}\right)\sim \left(\begin{array}{ccc} 1 & \ld & \ld^3
 \\
 \ld & 1 & \ld^2
 \\
 \ld^3 & \ld^2 &  1\end{array}\right), \label{eq:CKM}
\end{eqnarray}
with $\ld\approx 0.22$. Barring subtle cancelations, from the CKM matrix one may derive the following bounds for the mixing elements
\begin{eqnarray}\label{}
(U_u)_{c'u,u'c}, (U_d)_{s'd,d's}\lesssim\ld,~~
(U_u)_{c't,t'c}, (U_d)_{s'b,b's}\lesssim\ld^2,~~(U_u)_{t'u,u't}, (U_d)_{b'd,d'b}\lesssim\ld^3.
\label{eq:Uestimate}
\end{eqnarray}
A good pattern to achieve the CKM may be the saturation of the above inequalities (not all of them in each inequality but some combination). One may expect the similarity $W_q\sim U_q$ if the Yukawa couplings are symmetric in the sense of magnitude of order, otherwise they will differ significantly. In principle, $W_q$ can be made arbitrary, which has direct implications to the flavor changing processes. 
Note that for the down-type quark sector, $(U_d)_{s'd,d's}$ and $(U_d)_{b'd,d'b}$ are zero because of the mass matrix in Eq. \eqref{eq:quarkmass}.

In the doublet flavon model, the quark mass matrices turn out to be
\begin{eqnarray}\label{mass:singlet}
m_u^0 = \f{v_h}{\sqrt{2}}
\begin{pmatrix}
Y^u_{11}  & \wt Y^u_{12} \sqrt{r_{\mu\tau}} & \wt Y^u_{13}  \sqrt{r_{\mu\tau}}  \\
0  & Y^u_{22}  & Y^u_{23} \\
0 &  Y^u_{32}& Y^u_{33} \end{pmatrix}, ~~~ m_d^0 = \f{v_h}{\sqrt{2}}
\begin{pmatrix} Y^d_{11}  &  0  &  0 \\
 \wt Y^d_{21}  \sqrt{r_{\mu\tau}}    & Y^d_{22}  & Y^d_{23} \\
 \wt Y^d_{31}  \sqrt{r_{\mu\tau}}  &Y^d_{32} & Y^d_{33} \end{pmatrix}, \label{}
\end{eqnarray}
The resulting family mixings basically are the same as in the SM despite of the zeros. Notice that in principle one can turn off the $\wt {Y}_{i1}^d$-terms in Eq.~(\ref{F:doublet}), then recovering the Yukawa coupling structure similar to the singlet flavon model. However, it is unnatural to work in this limit without symmetry arguments~\footnote{For instance, in the supersymmetric models, such a limit is naturally realized by virtue of holomorphy.}. Later we will show that the generic Yukawa structure is strongly disfavored by the $(g-2)_\mu$ scenario.

\subsection{Flavor violating $Z'$ couplings}

In the singlet flavon model, the gauge sector induces quark FCNCs. It originates from the family non-universal local $U(1)_{B-L}$. Since $\Phi$ and ${\cal F}$ are singlet under the SM gauge symmetry, these fields doesn't contribute the mass terms for the SM gauge bosons and hence, there is no mixing between $Z'$ and SM gauge bosons at tree-level. As a result, flavor violating couplings arise only in the $Z'$ interactions. For the mass eigenstates of quarks, these couplings can be obtained as
\begin{eqnarray}\label{}
-{\cal L}^q_{Z'}&=&\f{g_{B-L}}{3}\bar q_{i}\gamma^\mu\left(V^q_{ij}-\gamma_5 A^q_{ij}\right) q_{j} Z'_{\mu}, \\
V^q_{ij}&=&\f{1}{2}\sum_{k=2,3}\left[(U_q^\dagger)_{ik}(U_q)_{kj}
+(W_q^\dagger)_{ik}(W_q)_{kj}\right]=\delta_{ij}-
\f{(U_q^\dagger)_{i1}(U_q)_{1j}
+(W_q^\dagger)_{i1}(W_q)_{1j}}{2},\cr A^q_{ij}&=&\f{1}{2}\sum_{k=2,3}\left[(U_q^\dagger)_{ik}(U_q)_{kj}
-(W_q^\dagger)_{ik}(W_q)_{kj}\right]=-\f{(U_q^\dagger)_{i1}(U_q)_{1j}
-(W_q^\dagger)_{i1}(W_q)_{1j}}{2}.
\end{eqnarray}
Therefore, the quark FCNCs in this model are related to $(1,i)$-elements of $U_q$ and $W_q$. Due to this fact, there are no quark FCNCs from down-type quarks since our model predicts $(U_d)_{1i} = 0 = (W_d)_{1i}$. In other words, in this model  within the down-type quark sector family mixing happens just within the second and third families, while $B-L$ charge assignment is universal in this subsector, and therefore the mass matrix and down-type quark $(B-L)_{\mu\tau}$ currents can be diagonalized simultaneously. By contrast, in the doublet flavon model generically speaking  $(U_d)_{1i},  (W_d)_{1i}\neq 0$, thus one expects full FCNCs.

\subsection{$\gamma-Z'$ kinetic mixing}

Till now we restrict our discussions to tree level, but at loop level there is an important correction to the properties of $Z'$, from the fields that are double charged under $(B-L)_{\mu\tau}$ and hypercharge $U(1)_Y$. They lead to kinematic mixing between the corresponding gauge fields. But we will simplify discussions by neglecting the effects from EWSB and assuming that $U(1)_{QED}$ along with $(B-L)_{\mu\tau}$ are our objects. Actually it gives the same expression as the calculation incorporating  EWSB~\cite{Kang:2010mh}. Then, the kinetic mixing term is 
\begin{equation}
\mathcal{L} \supset - \frac{\chi}{2} F^{\mu \nu} Z'_{\mu \nu},
\end{equation}
where $F_{\mu \nu}$ and $Z'_{\mu \nu}$ are the field strengths of the SM photon and $Z'$, respectively. In order to be ordinal canonical form, we redefine gauge fields as
\begin{equation}
A_{\mu} \rightarrow A_{\mu} - \frac{\chi}{\sqrt{1-\chi^2}} Z'_{\mu}, ~~~ Z'_{\mu} \rightarrow \frac{1}{\sqrt{1-\chi^2}} Z'_{\mu}.
\end{equation}
Therefore, fermions uncharged under  $(B-L)_{\mu\tau}$, e.g., the first generation of fermions also obtain $Z'$ couplings whose strength is $e Q_f \chi / \sqrt{1-\chi^2}$ where $e$ is the electric charge and $Q_f$ is the electromagnetic charge of fermion $f$. In particular, electron recouples to $Z'$ with a strength suppressed by loop. 

The mixing parameter $\chi$ can be calculated in analogy with the calculation of vacuum polarization diagram of gauge bosons \cite{Holdom:1985ag,Pich:1998xt}. In the $\overline{\rm MS}$ scheme, 
\begin{equation}
\chi = - \frac{e g_{B-L}}{12 \pi^2} \sum_f Q_f Q_f^{B-L} \left[ 6 \int_0^1 dx \, x(1-x) \ln \left( \frac{m_f^2 - k^2 x(1-x)}{\mu^2} \right) \right],
\end{equation}
where $Q_f^{B-L}$ is the $(B-L)_{\mu\tau}$ charge of fermion $f$ and $\mu$ is the renormalization scale. In the limit that the gauge boson momentum $k$ is much smaller than fermion masses, $\chi$ depends on $\ln (m_f^2 / \mu^2)$.
In the singlet flavon model, the second and third families of quarks and charged leptons and vector-like quarks contribute to $\chi$, and then
\begin{equation}
\chi = - \frac{e g_{B-L}}{18 \pi^2} \ln \left( \frac{m_c^2 m_t^2 m_{\mu}^3 m_{\tau}^3}{m_s m_b  M_U^8} \right).
\label{eq:kinmix}
\end{equation}
To get this expression we have assumed the boundary condition $\chi = 0$ at $M_U$. Note that in principle one can tune the boundary value to arrange a loop-tree cancellation, making the kinematic mixing at low energy smaller than the above estimation. In the doublet flavon model, on the other hand, there are no vector-like quarks, so that $\chi$ is smaller than that of the singlet flavon model because of the absence of $M_U$ contributions.

\section{$Z'$-related phenomenologies: $(g-2)_\mu$ confronting FCNCs }
\label{sec:ZpFCNC}

In this section we will first  study how the sub-GeV scale $(B-L)_{\mu\tau}$ gauge boson could  contribute positively and sizably to $(g-2)_\mu$, and then we study the flavor transitions induced by a very light $Z'$, which is relatively less studied in the new physics domain.

\subsection{$(g-2)_\mu$ from the light  $(B-L)_{\mu\tau}$ gauge boson }

The leptonic couplings of  $Z'$ in our model is independent to the quark flavor sector and is given by 
\begin{equation}
-\mathcal{L}_{Z'}^l = - g_{B-L} ( \bar{\mu} \gamma^{\mu} \mu + \bar{\tau} \gamma^{\mu} \tau+ \bar\nu_{\mu} \gamma^{\mu} \nu_{\mu}+ \bar\nu_{\tau} \gamma^{\mu} \nu_{\tau}) Z'_{\mu},
\end{equation}
with no couplings to electron-types at tree level. However, as discussed above, the electron couples to $Z'$ through the kinetic mixing. Therefore, when we explore the allowed parameter region for $(M_{Z'}, g_{B-L})$, we should take into account the bound related to $e-Z'$ coupling, like the Borexino bound \cite{Borexino1,Borexino2,Borexino3}.

The contribution from the light $Z'$ to the muon magnetic moment can be obtained as \cite{mug-2expression}~\footnote{If there are flavor violating couplings in charged lepton sector, like $g_{\mu\tau}^{L,R} \bar{\mu} \gamma^{\mu} P_{L,R} \tau Z'_{\mu}$, this expression should be modified by taking $g_{\mu\tau}^{L,R}$ contributions into account \cite{mug-2expression,Altmannshofer:2016brv}. In such a case, we can discuss lepton flavor violating processes and it is expected that there are some signatures in the future experiments \cite{Altmannshofer:2016brv,Foldenauer:2016rpi}.}
\begin{equation}
\Delta a_{\mu} = \frac{g_{B-L}^2}{8 \pi^2} \int_0^1 \frac{2 x^2 (1-x)}{x^2 + (M_{Z'}^2/m_{\mu}^2) (1-x)} dx.
\end{equation}
The current deviation $\Delta a_{\mu} \equiv a_{\mu}^{\rm exp} - a_{\mu}^{\rm SM}$ can be read as \cite{Hagiwara:2011af,Keshavarzi:2018mgv,Bennett:2006fi,Roberts:2010cj,Davier:2010nc,Davier:2017zfy,Davier:2019can}:
\begin{equation}
\Delta a_{\mu} = (27.06 \pm 7.26) \times 10^{-10}.
\label{eq:mug-2exp-SM}
\end{equation}

As for the leptonic side,  $Z'$ from the local  $(B-L)_{\mu\tau}$ model is similar to that from the $L_\mu-L_\tau$ model. So, as in the latter, the discrepancy of $(g-2)_\mu$ by virtue of  $Z'$ in our model is also constrained  from several aspects owing to the leptonic interactions of  $Z'$:
\begin{description}
\item[  Neutrino trident production] The region $M_{Z'}\gtrsim 400$ MeV~\cite{drident} accounting for $(g-2)_\mu$ has been clearly ruled out by  the neutrino trident production $\nu N\ra \nu \mu^+\mu^-$ which is searched by  CCFR~\cite{CCFR},

\item[ BaBar 4$\mu$ search] It searches the process $e^+e^-\ra \mu^+\mu^- Z'$ followed by $Z'\ra  \mu^+\mu^- $, and then further excludes the region $M_{Z'}\gtrsim 2m_\mu\simeq 210$ MeV \cite{BABAR}. 

\item[ BBN] The light $Z'$ would contribute to the effective relativistic degrees of freedom and spoils the BBN predictions. It yields the bound $M_{Z'}\gtrsim 1$ MeV~\cite{BBN1,BBN2,BBN3}. 
\end{description}
Therefore, only the narrow region $M_{Z'}\lesssim  2m_\mu$ survives. Light  $Z'$ in such a region has an immediate consequence to  $Z'$ decay: It overwhelmingly invisible decays into the neutrino pairs. Note that  despite of the coupling to quarks in our model,  $Z'$ is not subject to additional strong bounds.

\begin{figure}[htb]
  \begin{center}
    \includegraphics[width=0.6\textwidth,bb= 0 0 420 390]{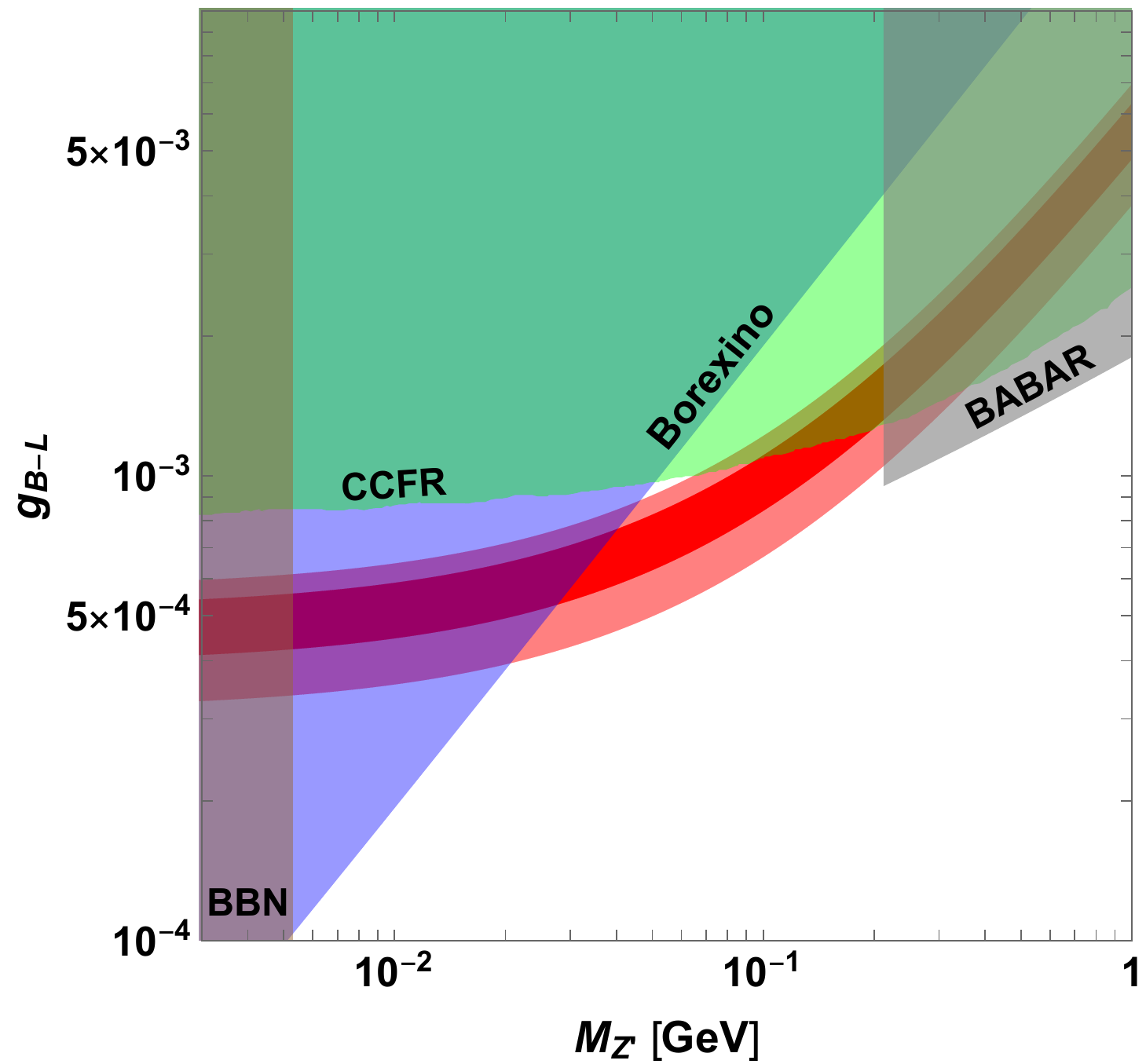}
    \caption{Allowed parameter space on the $M_{Z'}-g_{B-L}$ plane with relevant constraints. The favored regions for $\Delta a_{\mu} \equiv a_{\mu}^{\rm exp} - a_{\mu}^{\rm SM}$ from Eq. \eqref{eq:mug-2exp-SM} are shown by the red ($1\sigma$) and pink ($2\sigma$) bands. The other shaded regions show the constraints from CCFR (green) \cite{CCFR}, BaBar (gray) \cite{BABAR}, BBN (brown) \cite{BBN1,BBN2,BBN3} and Borexino with $M_U = 1$ TeV (blue) \cite{Borexino1,Borexino2,Borexino3}.}
    \label{fig:mug-2}
  \end{center}
\end{figure}

Using these results, we display the allowed parameter space on the $M_{Z'}-g_{B-L}$ plane in Fig. \ref{fig:mug-2}. 
The red and pink bands show the favored regions to explain $\Delta a_{\mu}$ within $1\sigma$ and $2\sigma$ level. Other shaded regions are constraints from CCFR (green), BaBar (gray) and BBN (brown).  We also show the constraint from the Borexino \cite{Borexino1,Borexino2,Borexino3} with blue shaded region by setting $M_U = 1$ TeV.
Therefore, the favored region is narrow for the value of $g_{B-L}$:
\begin{align}\label{10MeV}
&5.4 (4.3) \times 10^{-4} \leq |g_{B-L}| \leq 7.1 (7.8) \times 10^{-4} && (\text{for $M_{Z'} = 30$ MeV}),\\
&8.4 (6.7) \times 10^{-4} \leq |g_{B-L}| \leq 1.1 (1.2) \times 10^{-3} && (\text{for $M_{Z'} = 100$ MeV}),\label{100MeV}
\end{align}
from $1\sigma$ ($2\sigma$) of $\Delta a_{\mu}$. However, for $M_{Z'} = 30$ MeV, the Borexino constrains the upper region of coupling and gives $g_{B-L} \lesssim 5.5 \times 10^{-4}$. For $M_{Z'} = 100$ MeV, on the other hand, there is constraint from the CCFR which is  $g_{B-L} < 1.1 \times 10^{-3}$. 
In the discussion about the flavor physics in quark sector, we set $M_{Z'} = 30$ MeV and $g_{B-L} = 5.5 \times 10^{-4}$ as an example.

\subsection{Quark flavor violation from the gauge sectors}

\begin{figure}[htb]
  \begin{center}
    \includegraphics[width=0.6\textwidth,bb= 0 0 420 284]{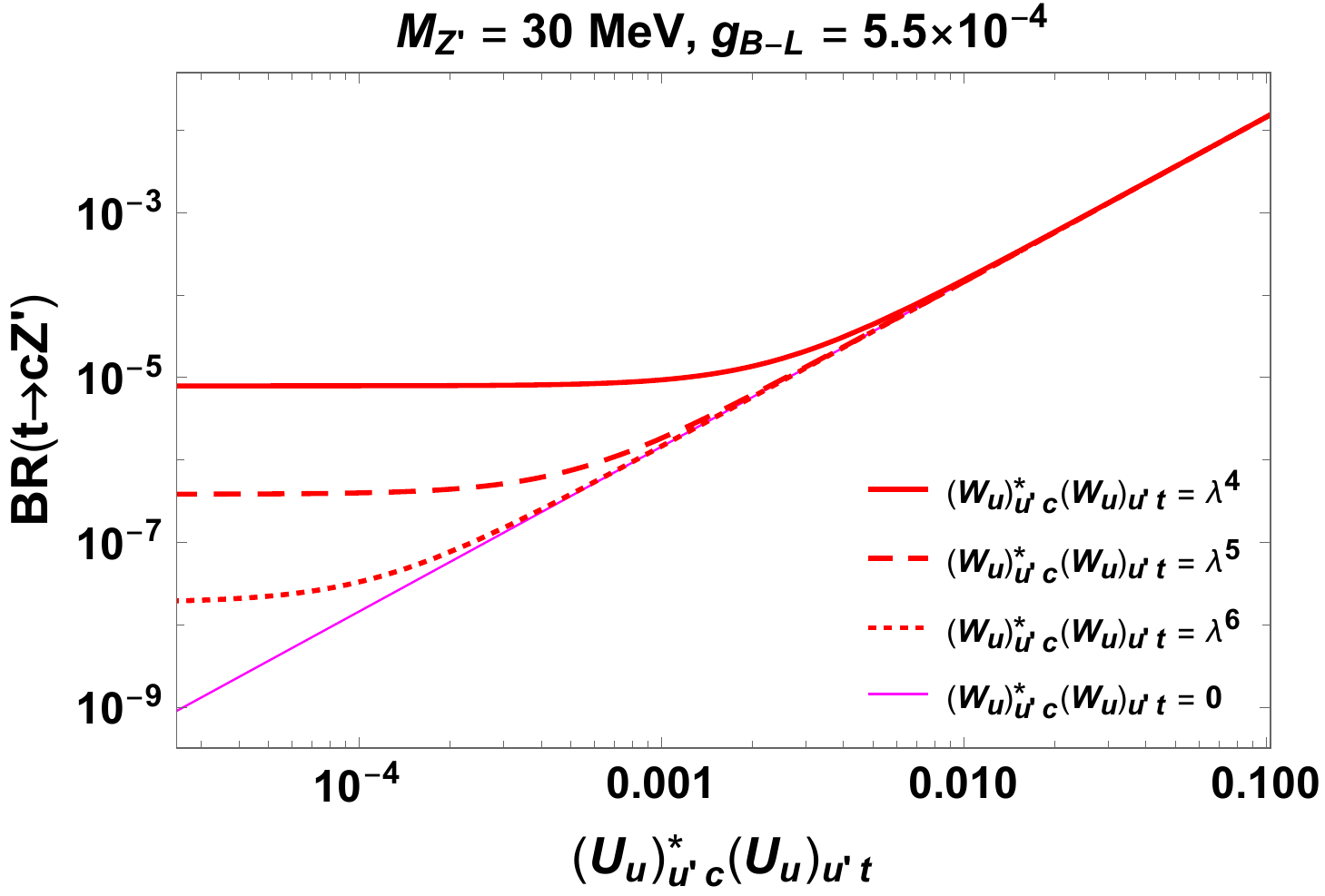}
    \caption{Our predictions of BR$(t \to c Z')$ as a function of $(U_u)^{\ast}_{u'c} (U_u)_{u't}$. The red solid, dashed and dotted lines show the case of $(W_u)^{\ast}_{u'c} (W_u)_{u't} = \lambda^4$, $\lambda^5$ and $\lambda^6$ with $\lambda = 0.22$, respectively. The magenta line shows the $(W_u)^{\ast}_{u'c} (W_u)_{u't} = 0$ case.}
    \label{fig:BRttocZp}
  \end{center}
\end{figure}

As explained above, we have flavor changing $Z'$ couplings in the up-type quark sector. One of the interesting processes is the flavor changing top quark decays: $t \to q Z'$ where $q = u, c$. We can calculate its decay width as \cite{Goodsell:2017pdq}
\begin{align}
\Gamma (t \to q Z') = \frac{m_t}{32 \pi} \lambda (1, x_q, x')^{1/2} \left[ \left( 1 + x_q - 2 x' + \frac{(1-x_q)^2}{x'} \right) (|(g_L^u)_{qt}|^2 + |(g_R^u)_{qt}|^2) \right. \nonumber \\
\biggl. - 12 \sqrt{x_q} {\rm Re} \left( (g_L^u)_{qt} (g_R^u)_{qt}^{\ast} \right) \biggr], \label{eq:ttocZp}
\end{align}
where $x_q \equiv m_q^2/m_t^2$, $x' \equiv M_{Z'}^2/m_t^2$ and $\lambda (x, y, z) = x^4 + y^4 + z^4 - 2 x^2 y^2 - 2 y^2 z^2 - 2 z^2 x^2$. $(g_L^u)_{qt}$ and $(g_R^u)_{qt}$ are our flavor violating $Z'$ coupling, defined as
\begin{align}
(g_L^u)_{ij} &= \frac{g_{B-L}}{3} (V_{ij}^u + A_{ij}^u) = -\frac{g_{B-L}}{3} (U_u^\dagger)_{i1} (U_u)_{1j}, \label{eq:guL} \\
(g_R^u)_{ij} &= \frac{g_{B-L}}{3} (V_{ij}^u - A_{ij}^u) = -\frac{g_{B-L}}{3} (W_u^\dagger)_{i1} (W_u)_{1j}. \label{eq:guR}
\end{align}
The branching ratio can be estimated by taking the ratio with the decay rate of $t \to b W$. Figure \ref{fig:BRttocZp} shows our prediction of BR$(t \to c Z')$ as a function of $(U_u)^{\ast}_{u'c} (U_u)_{u't}$.
The red solid, dashed and dotted lines show the case of $(W_u)^{\ast}_{u'c} (W_u)_{u't} = \lambda^4$, $\lambda^5$ and $\lambda^6$ with $\lambda = 0.22$, respectively, while the magenta line shows the result of $(W_u)^{\ast}_{u'c} (W_u)_{u't} = 0$. For this figure, we set $M_{Z'} = 30$ MeV and $g_{B-L} = 5.5 \times 10^{-4}$ for the explanation of $\Delta a_{\mu}$. Due to the smallness of the kinetic mixing $\chi$, the decay of $Z'$ is dominated by the decay $Z' \to \nu \bar \nu$ if $M_{Z'} = \mathcal{O}(10)$ MeV. Therefore, there are no clear bounds on BR$(t \to c Z')$ where $Z'$ decays to $\nu \bar{\nu}$. However, the decay width of top quark is dominated by the decay rate of $t \to b W$, and hence, BR$(t \to c Z')$ cannot be large. If we assume BR$(t \to c Z') < 10^{-4}$, $(U_u)^{\ast}_{u'c} (U_u)_{u't}$ should be smaller than $8 \times 10^{-3} \sim \lambda^{3.2}$ as long as $(W_u)^{\ast}_{u'c} (W_u)_{u't} < \lambda^4$. This is consistent with the bounds from the CKM matrix as shown in Eq. \eqref{eq:Uestimate}. This kind of rare decay, namely $t \to c + $missing energy, is studied in Ref. \cite{Abramowicz:2018rjq}, in the case that the top quark decays into a charm quark and a heavy stable particle. The expected limit on the branching ratio is $\mathcal{O}(10^{-4})$ which is calculated from 1.0 ${\rm ab}^{-1}$ collected at 380 GeV CLIC. Although this bound cannot be applied directly to our model, it is expected that we will obtain a clear bound on BR$(t \to c Z')$ at the future experiments. Therefore, if there would be some signals, we could explicitly constrain the elements of $U_u$ and $W_u$ and discuss the relations among these and other parameters. On the other hand, if there would be no signal and more severe bound would be obtained, we might be able to conclude that our model is inconsistent with the CKM matrix. Note that since the dominant contribution to BR$(t \to c Z')$ is given by $|(U_u)^{\ast}_{u'c} (U_u)_{u't}|^2 + |(W_u)^{\ast}_{u'c} (W_u)_{u't}|^2$ (see Eq.~\eqref{eq:roughwidth} below), we cannot use some cancellation between $(U_u)^{\ast}_{u'c} (U_u)_{u't}$ and $(W_u)^{\ast}_{u'c} (W_u)_{u't}$ in order to suppress BR$(t \to c Z')$.

We have two comments on this decay. First, if $M_{Z'}$ is small enough to be $x' \ll 1$, $\Gamma (t \to q Z')$ is approximately estimated as
\begin{align}
\Gamma (t \to q Z') &\approx \frac{m_t}{32 \pi} \lambda (1, x_q, x')^{1/2} \frac{1}{x'} (|(g_L^u)_{qt}|^2 + |(g_R^u)_{qt}|^2) \nonumber \\
&= \frac{m_t^3}{32 \pi} \lambda (1, x_q, x')^{1/2} \frac{|(g_L^u)_{qt}|^2 + |(g_R^u)_{qt}|^2}{M_{Z'}^2}. \label{eq:roughwidth}
\end{align}
Therefore, the prediction is proportional to $|g_{B-L}|^2/M_{Z'}^2$. This approximation can be valid for $M_{Z'} < \mathcal{O}(10)$ GeV. Second, the prediction of BR$(t \to u Z')$ is similar to BR$(t \to c Z')$ since the difference of decay width only comes from $x_q$ in Eq. \eqref{eq:ttocZp} and this is irrelevant to $\Gamma (t \to q Z')$. Therefore, we can estimate the prediction of BR$(t \to u Z')$ from Fig. \ref{fig:BRttocZp}. As a result, we obtain $(U_u)^{\ast}_{u'u} (U_u)_{u't} < 8 \times 10^{-3} \sim \lambda^{3.2}$ by assuming BR$(t \to u Z') < 10^{-4}$ and $(W_u)^{\ast}_{u'c} (W_u)_{u't} < \lambda^4$. This also gives reasonable value for $(U_u)_{u't} \lesssim \lambda^3$ when $(U_u)_{u'u} \sim 1$.

We would like to emphasize that we also obtain the same results for $t \to q Z'$ if we introduce extra $SU(2)_L$ Higgs doublet $H_{\mu\tau}$ instead of ${\cal F}$ and $U_{L, R}$. In this case, the down-type quarks also have flavor violating $Z'$ couplings as we mentioned above. Hence, the mixing angles in down-type quark sector may be severely constrained by $K$ and $B$ physics. We will show the results in this case and discuss how the constraints are severe.

\subsection{Severe constraint from $K$ and $B$ physics in the model with $H_{\mu\tau}$}

Since we consider the light $Z'$ mass in this model, it is important to search the constraint from meson decay processes. In this paper, we investigate the following decay processes: $B \to K \nu \bar{\nu}$ and $K \to \pi \nu \bar{\nu}$. Note that a light $Z'$ can be produced through $B$ or $K$ decays directly, and then the $Z'$ decays two neutrinos. Therefore, the branching ratios of these processes can be calculated as BR$(B \to K Z') \times$ BR$(Z' \to \nu \bar{\nu})$ and BR$(K \to \pi Z') \times$ BR$(Z' \to \nu \bar{\nu})$. In our setup, BR$(Z' \to \nu \bar{\nu}) \approx 1$ and
\begin{align}
{\rm BR}(B^+ \to K^+ Z') &= \frac{|(g_L^d)_{sb} + (g_R^d)_{sb}|^2}{64 \pi} \frac{\lambda (m_{B^+}, m_{K^+}, M_{Z'})^{3/2}}{M_{Z'}^2 m_{B^+}^3 \Gamma_{B^+}} \left[ f_+^{B^+ K^+}(M_{Z'}^2) \right]^2, \\
{\rm BR}(K^+ \to \pi^+ Z') &= \frac{|(g_L^d)_{ds} + (g_R^d)_{ds}|^2}{64 \pi} \frac{\lambda (m_{K^+}, m_{\pi^+}, M_{Z'})^{3/2}}{M_{Z'}^2 m_{K^+}^3 \Gamma_{K^+}} \left[ f_+^{K^+ \pi^+}(M_{Z'}^2) \right]^2, \\
{\rm BR}(K_L \to \pi^0 Z') &= \frac{|{\rm Im}(g_L^d)_{ds} + {\rm Im}(g_R^d)_{ds}|^2}{64 \pi} \frac{\lambda (m_{K_L}, m_{\pi^0}, M_{Z'})^{3/2}}{M_{Z'}^2 m_{K_L}^3 \Gamma_{K_L}} \left[ f_+^{K^0 \pi^0}(M_{Z'}^2) \right]^2,
\end{align}
where $m_X$ and $\Gamma_X$ are the mass and decay width of $X$ meson and $f_+^{M_1 M_2}(M_{Z'}^2)$ is $M_1 \to M_2$ form factor at $M_{Z'}^2$ \cite{Ball:2004ye,Mescia:2007kn}. In table \ref{tab:inputs}, we summarize the input parameters we used.
\begin{table}[h]
\begin{center}
\begin{tabular}{|c|c||c|c|}
\hline
$m_{K^+}$ & 493.677(16) MeV & $\tau_{K^+}$ & $1.2380(20) \times 10^{-8}$ s \\
$m_{K_L}$ & 497.611(13) MeV & $\tau_{K_L}$ & $5.116(21) \times 10^{-8}$ s \\
$m_{B^+}$ & 5279.32(14) MeV & $\tau_{B^+}$ & $1.638(4) \times 10^{-12}$ s \\
$m_{\pi^+}$ & 139.57061(24) MeV & & \\
$m_{\pi^0}$ & 134.9770(5) MeV & & \\
\hline
\end{tabular}
\caption{The input parameters we used \cite{PDG}.}
\label{tab:inputs}
\end{center}
\end{table}
The flavor violating $Z'$ couplings $(g_{L, R}^d)_{ij}$ can be written by
\begin{align}
(g_L^d)_{ij} &= \frac{g_{B-L}}{3} (V_{ij}^d + A_{ij}^d) = -\frac{g_{B-L}}{3} (U_d^\dagger)_{i1} (U_d)_{1j}, \label{eq:gdL} \\
(g_R^d)_{ij} &= \frac{g_{B-L}}{3} (V_{ij}^d - A_{ij}^d) = -\frac{g_{B-L}}{3} (W_d^\dagger)_{i1} (W_d)_{1j}. \label{eq:gdR}
\end{align}
Current experimental results and bounds for these branching ratios can be found in Refs. \cite{Lees:2013kla,Artamonov:2009sz,Ahn:2018mvc} as
\begin{align}
\Delta {\rm BR}(B^+ \to K^+ \nu \bar{\nu})_{\rm exp} &= (0.35^{+0.60}_{-0.15}) \times 10^{-5}, \label{eq:Bpexp} \\
{\rm BR}(K^+ \to \pi^+ Z')_{\rm exp} &< \mathcal{O}(10^{-10})\quad (\text{when}~M_{Z'} \simeq \mathcal{O}(10)~{\rm MeV}) \label{eq:Kpexp} \\
{\rm BR}(K_L \to \pi^0 \nu \bar{\nu})_{\rm exp} &< 3.0 \times 10^{-9}. \label{eq:KLexp}
\end{align}
Note that $\Delta {\rm BR}(B^+ \to K^+ \nu \bar{\nu})$ is the partial branching fractions defined in Ref. \cite{Lees:2013kla}. For BR$(K^+ \to \pi^+ Z')_{\rm exp}$, one can find the bound depending on $M_{Z'}$ from Fig. 18 of Ref. \cite{Artamonov:2009sz}. In this paper, we use the 90 \% C.L. upper limit: BR$(K^+ \to \pi^+ Z') < 1.0 \times 10^{-10}$ for $M_{Z'} = 30$ MeV. By using these, we discuss the bounds for the flavor violating $Z'$ couplings, $(g_{L, R}^d)_{ij}$.

\begin{figure}[htb]
  \begin{center}
    \includegraphics[width=0.55\textwidth,bb= 0 0 420 280]{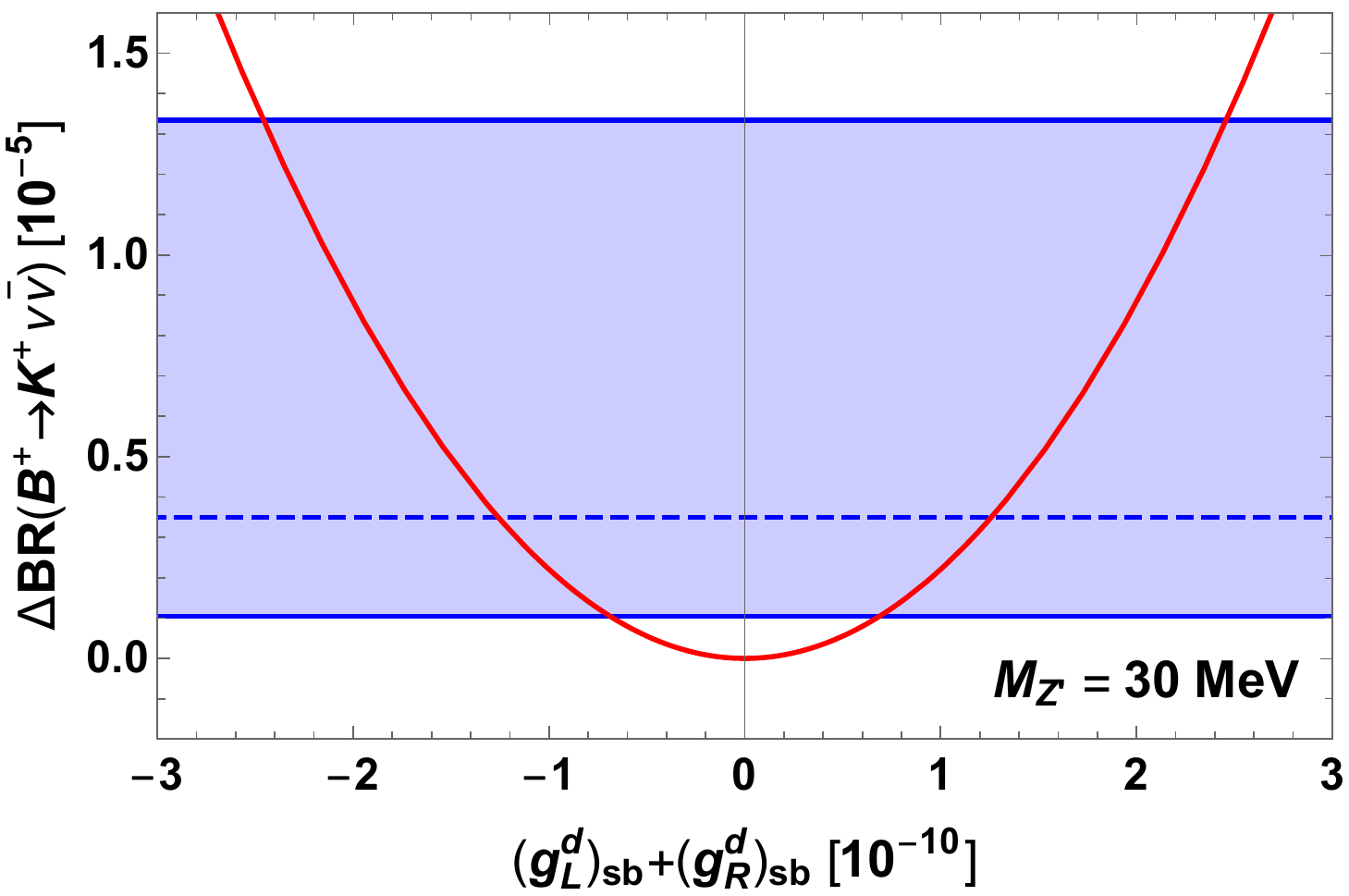}
    \caption{The red curve shows our prediction of $\Delta$BR$(B^+ \to K^+ \nu \bar{\nu})$ with $M_{Z'} = 30$ MeV. The blue dashed line is the central value of the experimental result Eq. \eqref{eq:Bpexp} and the shaded area shows $1.64\sigma$ allowed region from $\Delta$BR$(B^+ \to K^+ \nu \bar{\nu})_{\rm exp}$.}
    \label{fig:BptoKpnunu}
  \end{center}
\end{figure}

At first, we will show the bound from $B^+ \to K^+ \nu \bar{\nu}$ processes in Fig. \ref{fig:BptoKpnunu}. The horizontal axis shows the sum of the flavor violating $Z'$ couplings, $(g_L^d)_{sb} + (g_R^d)_{sb}$ and the vertical one is the branching ratio $\Delta$BR$(B^+ \to K^+ \nu \bar{\nu})$. The blue dashed line is the central value of the experimental result Eq. \eqref{eq:Bpexp} and the shaded area shows $1.64\sigma$ allowed region which corresponds to the 90 \% C.L.. The red curve is our prediction in the case of $M_{Z'} = 30$ MeV.
If we assume the new physics contribution are within $1.64\sigma$ region, the couplings should be satisfy
\begin{align}
0.69 \times 10^{-10} \leq | (g_L^d)_{sb} + (g_R^d)_{sb} | \leq 2.5 \times 10^{-10}.
\label{eq:boundBp}
\end{align}
Note that from Eqs. \eqref{eq:gdL} and \eqref{eq:gdR}, $(g_L^d)_{sb}$ is related with the CKM matrix elements as shown in Eq. \eqref{eq:CKM}, while $(g_R^d)_{sb}$ is not related with any SM observables. Therefore, we can choose any values to explain the relation between $(g_L^d)_{sb}$ and the CKM matrix. If we set $(g_R^d)_{sb} = 0$ and $g_{B-L} = 5.5 \times 10^{-4}$, we obtain $3.7 \times 10^{-7} \leq | (U_d)^\ast_{d's} (U_d)_{d'b} | \leq 1.3 \times 10^{-6}$. This means that $| (U_d)^\ast_{d's} (U_d)_{d'b} | \sim \lambda^9$ and then, this bound is severe compared with bounds from the CKM matrix as shown in Eq. \eqref{eq:Uestimate}. If one considers $(g_R^d)_{sb} \neq 0$ and permits the cancellation between $(g_L^d)_{sb}$ and $(g_R^d)_{sb}$, one can obtain the reasonable values for $(U_d)_{d's}$ and $(U_d)_{d'b}$. Note that the magnitude of the bound for $| (U_d)^\ast_{d's} (U_d)_{d'b} |$ is not so much dependent on $g_{B-L}$ as long as we consider $\Delta a_{\mu}$ bound. The bound with different values of $g_{B-L}$ can be obtained by multiplying $(\frac{5.5 \times 10^{-4}}{g_{B-L}})$.

\begin{figure}[htb]
  \begin{center}
    \includegraphics[width=0.445\textwidth,bb= 0 0 420 280]{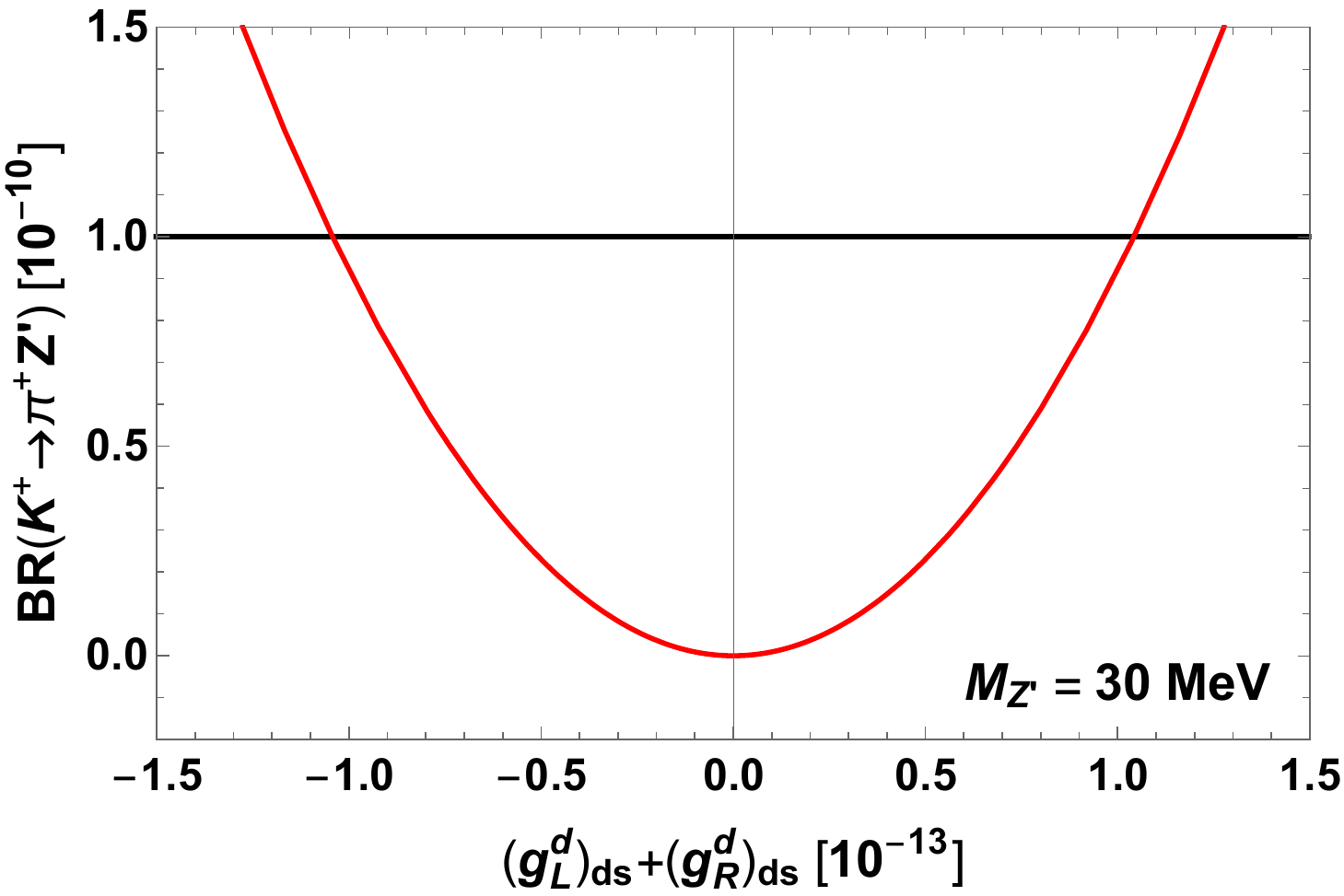}\hspace{1.0em}
    \includegraphics[width=0.43\textwidth,bb= 0 0 420 290]{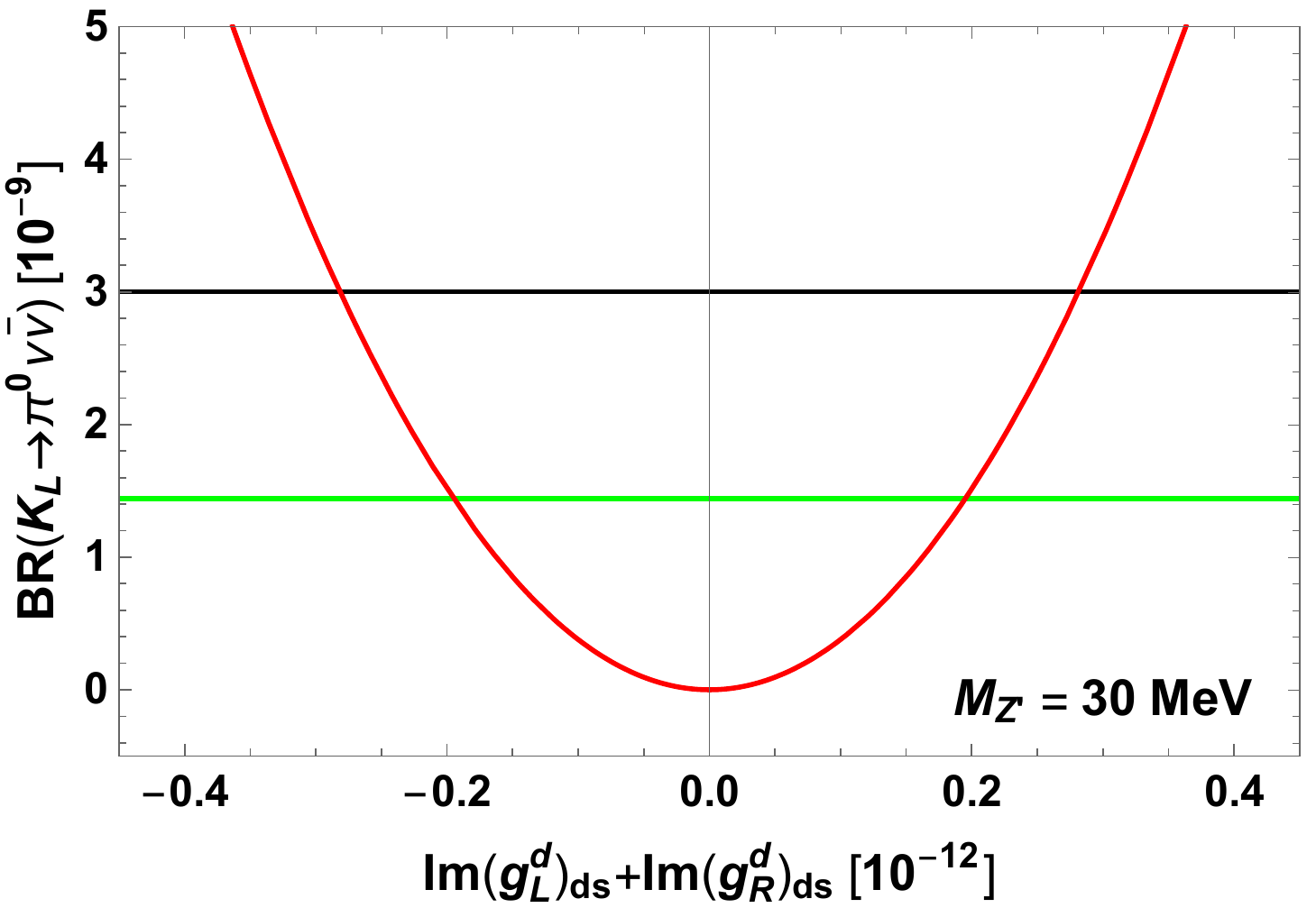}
    \caption{The red curve in each panel shows our prediction of BR$(K^+ \to \pi^+ Z')$ (left) and BR$(K_L \to \pi^0 \nu \bar{\nu})$ (right) with $M_{Z'} = 30$ MeV. The black solid lines in each panel show the upper bound from BR$(K^+ \to \pi^+ Z')_{\rm exp}$ (Eq. \eqref{eq:Kpexp}) and BR$(K_L \to \pi^0 \nu \bar{\nu})_{\rm exp}$ (Eq. \eqref{eq:KLexp}). In the right panel, the green line shows the Grossman-Nir bound \cite{Grossman:1997sk}.}
    \label{fig:Ktopinunu}
  \end{center}
\end{figure}

Next, we find bounds on $(g_{L, R}^d)_{ds}$ from $K^+$ and $K_L$ decays. The predictions of our model are shown in Fig. \ref{fig:Ktopinunu} as the red curves. In the left panel, the horizontal and vertical axes show the sum of the flavor violating $Z'$ couplings, $(g_L^d)_{ds} + (g_R^d)_{ds}$ and BR$(K^+ \to \pi^+ Z')$, while in the right panel, those show the sum of imaginary part of the flavor violating $Z'$ couplings, ${\rm Im}(g_L^d)_{ds} + {\rm Im}(g_R^d)_{ds}$ and BR$(K_L \to \pi^0 \nu \bar{\nu})$. The solid black lines in each panel show the upper bound from BR$(K^+ \to \pi^+ Z')_{\rm exp}$ (Eq. \eqref{eq:Kpexp}) and BR$(K_L \to \pi^0 \nu \bar{\nu})_{\rm exp}$ (Eq. \eqref{eq:KLexp}). In the right panel, we also show the Grossman-Nir bound \cite{Grossman:1997sk} as the green line. This bound can be calculated by
\begin{align}
{\rm BR}(K_L \to \pi^0 \nu \bar{\nu}) < 4.3 \times {\rm BR}(K^+ \to \pi^+ \nu \bar{\nu}),
\end{align}
where the factor $4.3$ comes from the ratio of the lifetimes of $K^+$ and $K_L$ and isospin breaking effect. In this figure, we use $90 \%$ C.L. upper bound of BR$(K^+ \to \pi^+ \nu \bar{\nu}) < 3.35 \times 10^{-10}$ from Ref. \cite{Artamonov:2009sz}.
From this figure, we can extract the bounds on $(g_{L, R}^d)_{ds}$ as
\begin{align}
|(g_L^d)_{ds} + (g_R^d)_{ds}| &< 0.11 \times 10^{-12}, \\
 |{\rm Im}(g_L^d)_{ds} + {\rm Im}(g_R^d)_{ds}| &< 0.28 \times 10^{-12} ~~ (0.20 \times 10^{-12} ~\text{(GN bound)}).
\label{eq:boundK}
\end{align}
Therefore, we found that BR$(K^+ \to \pi^+ Z')$ gives stronger constraint than BR$(K_L \to \pi^0 \nu \bar{\nu})$. When we set $(g_R^d)_{ds} = 0$ and $g_{B-L} = 5.5 \times 10^{-4}$, this bound become $|(U_d)^\ast_{d'd} (U_d)_{d's}| < 6.2 \times 10^{-10}$ which gives also severe bound on $U_d$. Obviously, this bound can be weakened by choosing appropriate values of $(g_R^d)_{ds}$.

Note that by combining the constraints from $B^+ \to K^+ \nu \bar{\nu}$ and $K^+ \to \pi^+ \nu \bar{\nu}$ decays with the unitary condition of $U_d$, we can estimate the size of each element of $(U_d)_{d'i} ~ (i = d, s, b)$ when $(g_R^d)_{ij} = 0$ and $g_{B-L} = 5.5 \times 10^{-4}$. The allowed patterns are only following two cases:
\begin{align}
\begin{array}{ccc}
~~|(U_d)_{d'd}|~~ & ~~|(U_d)_{d's}|~~ & ~~|(U_d)_{d'b}|~~ \\[0.5ex] \hline
< 10^{-9} & \sim 1 & \simeq \mathcal{O}(10^{-6}) \\
< 10^{-3} & \simeq \mathcal{O}(10^{-6}) & \sim 1
\end{array} \nonumber
\end{align}
Therefore, $|(U_d)_{d'd}|$ cannot be $\mathcal{O}(1)$ in this model unless we consider $(g_R^d)_{ij} \neq 0$. As a result, if we introduce $H_{\mu\tau}$ to the model, it is inconsistent with the proper structure of the CKM matrix.

\section{Conclusions and discussions}
\label{sec:summary}

In this paper, we discuss flavor violating processes involving $Z'$ from flavored $U(1)_{B-L}$, denoted as $({B-L})_{\mu\tau}$. Under this symmetry, the second and third families of fermions are charged and then, enough numbers of right-handed neutrinos to generate tiny neutrino masses are naturally introduced. Since we consider this new symmetry is broken at the scale far below the weak scale, the $Z'$ has small mass, $\mathcal{O}(10)$ MeV. In such a light $Z'$ case, it is favor to explain the deviation of the muon anomalous magnetic moment. In order to explain the current deviation, the size of $({B-L})_{\mu\tau}$ coupling $g_{B-L}$ need to be about $5.5 \times 10^{-4}$.

On the other hand, we should introduce the flavon field and vector-like quarks to accommodate the structure of the CKM matrix. Especially, we introduce the SM singlet flavon ${\cal F}$ and up-type vector-like quarks to the model. Due to this and flavor non-universal $({B-L})_{\mu\tau}$ charges, $Z'$ couplings of up-type quarks become the flavor violating ones, so that we investigate the predictions and constraints from quark flavor physics in this paper. One of the interesting processes is $t \to q Z'$ decay and we obtained the result that the mixing angles in the up-type quark sector is $(U_u)^{\ast}_{u'c} (U_u)_{u't} < 8 \times 10^{-3} \sim \lambda^{3.2}$ by assuming BR$(t \to c Z') < 10^{-4}$. Interestingly, this bound is consistent with the CKM matrix.

We also investigate the model with extra $SU(2)_L$ Higgs doublet as a flavon. In this case, no vector-like quarks are needed to accommodate the CKM matrix. However, the down-type quark sector also has the flavor violating $Z'$ couplings and therefore, strong constraints come from the $K$ and $B$ meson decays. As a result, we obtained the conclusion that $(U_d)_{d'i}$ for $i = d, s, b$ are severely constrained. The important point of this conclusion is that $(U_d)_{d'd}$ element cannot be $\mathcal{O}(1)$ and then, proper structure of the CKM matrix is failed unless we consider the unnatural cancellation in $(g_L^d)_{ds} + (g_R^d)_{ds}$~\footnote{Although such a cancellation can suppress $B\ra KZ'(\ra \nu\bar\nu)$, it is impossible to suppress the fully leptonic decay like $B_s\ra\mu\bar \mu$. However, our $Z'$ does not contribute to decays of pseudo-scalar meson to lepton pair since only vector current exists in the $Z'$ sector. Therefore, our model is free from such experimental results, including precise bound on $D \to \mu \mu$ \cite{Aaij:2013cza}.}. 

There are some on-going experiments to explore in light $Z'$ mass regions: for example NA64 with electron beam \cite{NA64e1,NA64e2,NA64e3}, NA64 with muon beam \cite{NA64m} and DUNE \cite{Altmannshofer:2019zhy}. Therefore, it can be expected that there are some signals in this regions. If no signal is observed, the explanation of $(g-2)_{\mu}$ with light $Z'$ will be excluded.

To end up this paper we would like to relate this work with a previous work, Ref.~\cite{Kaneta:2016vkq}. It discussed the scenario that the lightest right-handed neutrino plays the role of dark matter candidate in the minimal gauged $B-L$ model, where all three families of fermions are charged under it universally. It is nontrivial to carry out a similar study in our model because we merely have two RHNs, thus neither of them allowed to decouple from the left-handed leptons. As a result, the lighter RHN must be sufficiently light to be long-lived. Whether such a dark matter candidate is allowed is of interest. By the way, the $Z_2$ residue of $B-L$, which could provide the protecting symmetry for other dark matter extensions~\cite{Cai:2018nob}, may be lost owing to the presence of extra fields with VEVs but without properly assigned quantum numbers.

\section{Acknowledgements}

We thank the early collaborations with Wenyu Wang and Yang Xu. This work is supported in part by the National Science Foundation of China (11775086).

\appendix


\vspace{-.3cm}
\begin{thebibliography}{99}



\bibitem{Davidson:1978pm} 
  A.~Davidson,
  Phys.\ Rev.\ D {\bf 20} (1979) 776.


\bibitem{Mohapatra:1980qe} 
  R.~N.~Mohapatra and R.~E.~Marshak,
  Phys.\ Rev.\ Lett.\  {\bf 44} (1980) 1316
  [Erratum: Phys.\ Rev.\ Lett.\  {\bf 44} (1980) 1643].

\bibitem{Marshak:1979fm} 
  R.~E.~Marshak and R.~N.~Mohapatra,
  Phys.\ Lett.\  {\bf 91B} (1980) 222.

\bibitem{Seesaw1}
  P.~Minkowski,
  Phys.\ Lett.\  {\bf 67B} (1977) 421.

\bibitem{Seesaw2}
  M.~Gell-Mann, P.~Ramond and R.~Slansky,
  Conf.\ Proc.\ C {\bf 790927} (1979) 315
  [arXiv:1306.4669 [hep-th]].

\bibitem{Seesaw3}
  T.~Yanagida,
  Conf.\ Proc.\ C {\bf 7902131} (1979) 95.

\bibitem{Seesaw4}
  S.~L.~Glashow,
  NATO Sci.\ Ser.\ B {\bf 61} (1980) 687.

\bibitem{Seesaw5}
  R.~N.~Mohapatra and G.~Senjanovic,
  Phys.\ Rev.\ Lett.\  {\bf 44} (1980) 912.

\bibitem{Seesaw6}
  R.~N.~Mohapatra and G.~Senjanovic,
  Phys.\ Rev.\ D {\bf 23} (1981) 165.

\bibitem{Seesaw7}
  J.~Schechter and J.~W.~F.~Valle,
  Phys.\ Rev.\ D {\bf 22} (1980) 2227.

\bibitem{Aghanim:2018eyx} 
  N.~Aghanim {\it et al.} [Planck Collaboration],
  arXiv:1807.06209 [astro-ph.CO].

\bibitem{Esteban:2018azc} 
  I.~Esteban, M.~C.~Gonzalez-Garcia, A.~Hernandez-Cabezudo, M.~Maltoni and T.~Schwetz,
  JHEP {\bf 1901} (2019) 106
  [arXiv:1811.05487 [hep-ph]].


\bibitem{Hagiwara:2011af} 
  K.~Hagiwara, R.~Liao, A.~D.~Martin, D.~Nomura and T.~Teubner,
  J.\ Phys.\ G {\bf 38} (2011) 085003
  [arXiv:1105.3149 [hep-ph]].

\bibitem{Keshavarzi:2018mgv} 
  A.~Keshavarzi, D.~Nomura and T.~Teubner,
  Phys.\ Rev.\ D {\bf 97} (2018) 114025
  [arXiv:1802.02995 [hep-ph]].

\bibitem{Bennett:2006fi} 
  G.~W.~Bennett {\it et al.} [Muon g-2 Collaboration],
  Phys.\ Rev.\ D {\bf 73} (2006) 072003
  [hep-ex/0602035].

\bibitem{Davier:2010nc}
  M.~Davier, A.~Hoecker, B.~Malaescu and Z.~Zhang,
  Eur.\ Phys.\ J.\ C {\bf 71} (2011) 1515
  [Erratum: Eur.\ Phys.\ J.\ C {\bf 72} (2012) 1874]
  [arXiv:1010.4180 [hep-ph]].

\bibitem{Davier:2017zfy}
  M.~Davier, A.~Hoecker, B.~Malaescu and Z.~Zhang,
  Eur.\ Phys.\ J.\ C {\bf 77} (2017) 827
  [arXiv:1706.09436 [hep-ph]].

\bibitem{Davier:2019can}
  M.~Davier, A.~Hoecker, B.~Malaescu and Z.~Zhang,
  arXiv:1908.00921 [hep-ph].

\bibitem{Roberts:2010cj} 
  B.~L.~Roberts,
  Chin.\ Phys.\ C {\bf 34} (2010) 741
  [arXiv:1001.2898 [hep-ex]].

\bibitem{darkA}
  M.~Pospelov,
  Phys.\ Rev.\ D {\bf 80} (2009) 095002
  [arXiv:0811.1030 [hep-ph]].

 \bibitem{darkA:new}
  G.~Mohlabeng,
  Phys.\ Rev.\ D {\bf 99} (2019) 115001
  [arXiv:1902.05075 [hep-ph]].

\bibitem{Alonso:2017uky}
  R.~Alonso, P.~Cox, C.~Han and T.~T.~Yanagida,
  Phys.\ Lett.\ B {\bf 774} (2017) 643
  [arXiv:1705.03858 [hep-ph]].

\bibitem{Babu:2017olk} 
  K.~S.~Babu, A.~Friedland, P.~A.~N.~Machado and I.~Mocioiu,
  JHEP {\bf 1712} (2017) 096
  [arXiv:1705.01822 [hep-ph]].

\bibitem{Elahi:2019drj} 
  F.~Elahi and A.~Martin,
  Phys.\ Rev.\ D {\bf 100}, (2019) 035016
  [arXiv:1905.10106 [hep-ph]].

\bibitem{LmuLtau1}
  X.-G.~He, G.~C.~Joshi, H.~Lew and R.~R.~Volkas,
  Phys.\ Rev.\ D {\bf 43} (1991) 22.

\bibitem{LmuLtau2} 
  X.~G.~He, G.~C.~Joshi, H.~Lew and R.~R.~Volkas,
  Phys.\ Rev.\ D {\bf 44} (1991) 2118.

\bibitem{LmuLtau:g-2}
  S.~Baek, N.~G.~Deshpande, X.~G.~He and P.~Ko,
  Phys.\ Rev.\ D {\bf 64} (2001) 055006
  [hep-ph/0104141].


\bibitem{CCFR}
  S.~R.~Mishra {\it et al.} [CCFR Collaboration],
  Phys.\ Rev.\ Lett.\  {\bf 66} (1991) 3117.

\bibitem{BABAR}
  J.~P.~Lees {\it et al.} [BaBar Collaboration],
  Phys.\ Rev.\ D {\bf 94} (2016) 011102
  [arXiv:1606.03501 [hep-ex]].

\bibitem{BBN1}
  B.~Ahlgren, T.~Ohlsson and S.~Zhou,
  Phys.\ Rev.\ Lett.\  {\bf 111} (2013) 199001
  [arXiv:1309.0991 [hep-ph]].

\bibitem{BBN2}
  A.~Kamada and H.~B.~Yu,
  Phys.\ Rev.\ D {\bf 92} (2015) 113004
  [arXiv:1504.00711 [hep-ph]].

\bibitem{BBN3}
  M.~Escudero, D.~Hooper, G.~Krnjaic and M.~Pierre,
  JHEP {\bf 1903} (2019) 071
  [arXiv:1901.02010 [hep-ph]].

\bibitem{Bauer:2018onh} 
  M.~Bauer, P.~Foldenauer and J.~Jaeckel,
  JHEP {\bf 1807} (2018) 094
  [arXiv:1803.05466 [hep-ph]].

\bibitem{Davoudiasl:2014kua} 
  H.~Davoudiasl, H.~S.~Lee and W.~J.~Marciano,
  Phys.\ Rev.\ D {\bf 89} (2014) 095006
  [arXiv:1402.3620 [hep-ph]].

\bibitem{Fuyuto:2014cya} 
  K.~Fuyuto, W.~S.~Hou and M.~Kohda,
  Phys.\ Rev.\ Lett.\  {\bf 114} (2015) 171802
  [arXiv:1412.4397 [hep-ph]].

\bibitem{Jeong:2015bbi} 
  Y.~S.~Jeong, C.~S.~Kim and H.~S.~Lee,
  Int.\ J.\ Mod.\ Phys.\ A {\bf 31} (2016) 1650059
  [arXiv:1512.03179 [hep-ph]].

\bibitem{Fuyuto:2015gmk} 
  K.~Fuyuto, W.~S.~Hou and M.~Kohda,
  Phys.\ Rev.\ D {\bf 93} (2016) 054021
  [arXiv:1512.09026 [hep-ph]].

\bibitem{Kaneta:2016vkq} 
  K.~Kaneta, Z.~Kang and H.~S.~Lee,
  JHEP {\bf 1702} (2017) 031
  [arXiv:1606.09317 [hep-ph]].

\bibitem{Datta:2017pfz} 
  A.~Datta, J.~Liao and D.~Marfatia,
  Phys.\ Lett.\ B {\bf 768} (2017) 265
  [arXiv:1702.01099 [hep-ph]].

\bibitem{Araki:2017wyg} 
  T.~Araki, S.~Hoshino, T.~Ota, J.~Sato and T.~Shimomura,
  Phys.\ Rev.\ D {\bf 95} (2017) 055006
  [arXiv:1702.01497 [hep-ph]].

\bibitem{Gninenko:2018tlp} 
  S.~N.~Gninenko and N.~V.~Krasnikov,
  Phys.\ Lett.\ B {\bf 783} (2018) 24
  [arXiv:1801.10448 [hep-ph]].

\bibitem{Kamada:2018zxi} 
  A.~Kamada, K.~Kaneta, K.~Yanagi and H.~B.~Yu,
  JHEP {\bf 1806} (2018) 117
  [arXiv:1805.00651 [hep-ph]].

\bibitem{Biswas:2019twf}
  A.~Biswas and A.~Shaw,
  JHEP {\bf 1905} (2019) 165
  [arXiv:1903.08745 [hep-ph]].





\bibitem{Altmannshofer:2016jzy} 
  W.~Altmannshofer, S.~Gori, S.~Profumo and F.~S.~Queiroz,
  JHEP {\bf 1612} (2016) 106
  [arXiv:1609.04026 [hep-ph]].

\bibitem{Ko:2017quv} 
  P.~Ko, T.~Nomura and H.~Okada,
  Phys.\ Lett.\ B {\bf 772} (2017) 547
  [arXiv:1701.05788 [hep-ph]].

\bibitem{DiChiara:2017cjq} 
  S.~Di Chiara, A.~Fowlie, S.~Fraser, C.~Marzo, L.~Marzola, M.~Raidal and C.~Spethmann,
  Nucl.\ Phys.\ B {\bf 923} (2017) 245
  [arXiv:1704.06200 [hep-ph]].

\bibitem{Bonilla:2017lsq} 
  C.~Bonilla, T.~Modak, R.~Srivastava and J.~W.~F.~Valle,
  Phys.\ Rev.\ D {\bf 98} (2018) 095002
  [arXiv:1705.00915 [hep-ph]].

\bibitem{Bian:2017rpg} 
  L.~Bian, S.~M.~Choi, Y.~J.~Kang and H.~M.~Lee,
  Phys.\ Rev.\ D {\bf 96} (2017) 075038
  [arXiv:1707.04811 [hep-ph]].

\bibitem{Falkowski:2018dsl} 
  A.~Falkowski, S.~F.~King, E.~Perdomo and M.~Pierre,
  JHEP {\bf 1808} (2018) 061
  [arXiv:1803.04430 [hep-ph]].

\bibitem{Arcadi:2018tly} 
  G.~Arcadi, T.~Hugle and F.~S.~Queiroz,
  Phys.\ Lett.\ B {\bf 784} (2018) 151
  [arXiv:1803.05723 [hep-ph]].

\bibitem{Chun:2018ibr} 
  E.~J.~Chun, A.~Das, J.~Kim and J.~Kim,
  JHEP {\bf 1902} (2019) 093
  [arXiv:1811.04320 [hep-ph]].

\bibitem{Hutauruk:2019crc} 
  P.~T.~P.~Hutauruk, T.~Nomura, H.~Okada and Y.~Orikasa,
  Phys.\ Rev.\ D {\bf 99} (2019) 055041
  [arXiv:1901.03932 [hep-ph]].


\bibitem{Berezhiani:1983hm} 
  Z.~G.~Berezhiani,
  Phys.\ Lett.\  {\bf 129B} (1983) 99.

\bibitem{Chang:1986bp} 
  D.~Chang and R.~N.~Mohapatra,
  Phys.\ Rev.\ Lett.\  {\bf 58} (1987) 1600.

\bibitem{Rajpoot:1987fca} 
  S.~Rajpoot,
  Mod.\ Phys.\ Lett.\ A {\bf 2}  (1987) 307
  [Erratum: Mod.\ Phys.\ Lett.\ A {\bf 2} (1987) 541].

\bibitem{Rajpoot:1986nv} 
  S.~Rajpoot,
  Phys.\ Lett.\ B {\bf 191} (1987) 122.

\bibitem{Rajpoot:1987ji} 
  S.~Rajpoot,
  Phys.\ Rev.\ D {\bf 36} (1987) 1479.

\bibitem{Davidson:1987mh} 
  A.~Davidson and K.~C.~Wali,
  Phys.\ Rev.\ Lett.\  {\bf 59} (1987) 393.

\bibitem{Rajpoot:1988gx} 
  S.~Rajpoot,
  Phys.\ Rev.\ D {\bf 39} (1989) 351.

\bibitem{Berezhiani:1991ds} 
  Z.~G.~Berezhiani and R.~Rattazzi,
  Phys.\ Lett.\ B {\bf 279} (1992) 124.


\bibitem{Kang:2010mh} 
  Z.~Kang, T.~Li, T.~Liu, C.~Tong and J.~M.~Yang,
  JCAP {\bf 1101} (2011) 028
  [arXiv:1008.5243 [hep-ph]].


\bibitem{Holdom:1985ag} 
  B.~Holdom,
  Phys.\ Lett.\  {\bf 166B} (1986) 196.

\bibitem{Pich:1998xt} 
  A.~Pich,
  hep-ph/9806303.






\bibitem{Borexino1}
  G.~Bellini {\it et al.},
  Phys.\ Rev.\ Lett.\  {\bf 107} (2011) 141302
  [arXiv:1104.1816 [hep-ex]].

\bibitem{Borexino2}
  R.~Harnik, J.~Kopp and P.~A.~N.~Machado,
  JCAP {\bf 1207} (2012) 026
  [arXiv:1202.6073 [hep-ph]].

\bibitem{Borexino3}
  M.~Agostini {\it et al.} [Borexino Collaboration],
  Phys.\ Rev.\ D {\bf 100} (2019) 082004
  [arXiv:1707.09279 [hep-ex]].

\bibitem{mug-2expression}
  J.~P.~Leveille,
  Nucl.\ Phys.\ B {\bf 137} (1978) 63.

\bibitem{Altmannshofer:2016brv} 
  W.~Altmannshofer, C.~Y.~Chen, P.~S.~Bhupal Dev and A.~Soni,
  Phys.\ Lett.\ B {\bf 762} (2016) 389
  [arXiv:1607.06832 [hep-ph]].

\bibitem{Foldenauer:2016rpi} 
  P.~Foldenauer and J.~Jaeckel,
  JHEP {\bf 1705} (2017) 010
  [arXiv:1612.07789 [hep-ph]].


\bibitem{drident}
  W.~Altmannshofer, S.~Gori, M.~Pospelov and I.~Yavin,
  Phys.\ Rev.\ Lett.\  {\bf 113} (2014) 091801. 
  



  
  
\bibitem{Goodsell:2017pdq} 
  M.~D.~Goodsell, S.~Liebler and F.~Staub,
  Eur.\ Phys.\ J.\ C {\bf 77} (2017) 758
  [arXiv:1703.09237 [hep-ph]].
  
  

\bibitem{Abramowicz:2018rjq}
  H.~Abramowicz {\it et al.} [CLICdp Collaboration],
  arXiv:1807.02441 [hep-ex].

  
\bibitem{Ball:2004ye} 
  P.~Ball and R.~Zwicky,
  Phys.\ Rev.\ D {\bf 71} (2005) 014015
  [hep-ph/0406232].
  

\bibitem{Mescia:2007kn} 
  F.~Mescia and C.~Smith,
  Phys.\ Rev.\ D {\bf 76} (2007) 034017
  [arXiv:0705.2025 [hep-ph]].

\bibitem{PDG}
  M.~Tanabashi {\it et al.} [Particle Data Group],
  Phys.\ Rev.\ D {\bf 98} (2018) 030001.

\bibitem{Lees:2013kla} 
  J.~P.~Lees {\it et al.} [BaBar Collaboration],
  Phys.\ Rev.\ D {\bf 87} (2013) 112005
  [arXiv:1303.7465 [hep-ex]].

\bibitem{Artamonov:2009sz} 
  A.~V.~Artamonov {\it et al.} [BNL-E949 Collaboration],
  Phys.\ Rev.\ D {\bf 79} (2009) 092004
  [arXiv:0903.0030 [hep-ex]].

\bibitem{Ahn:2018mvc} 
  J.~K.~Ahn {\it et al.} [KOTO Collaboration],
  Phys.\ Rev.\ Lett.\  {\bf 122} (2019) 021802
  [arXiv:1810.09655 [hep-ex]].

\bibitem{Grossman:1997sk} 
  Y.~Grossman and Y.~Nir,
  Phys.\ Lett.\ B {\bf 398} (1997) 163
  [hep-ph/9701313].
  
\bibitem{Aaij:2013cza} 
  R.~Aaij {\it et al.} [LHCb Collaboration],
  Phys.\ Lett.\ B {\bf 725}, 15 (2013)
  [arXiv:1305.5059 [hep-ex]].

\bibitem{NA64e1}
  S.~Andreas {\it et al.},
  arXiv:1312.3309 [hep-ex].

\bibitem{NA64e2}
  D.~Banerjee {\it et al.} [NA64 Collaboration],
  Phys.\ Rev.\ Lett.\  {\bf 118} (2017) 011802
  [arXiv:1610.02988 [hep-ex]].

\bibitem{NA64e3}
  D.~Banerjee {\it et al.} [NA64 Collaboration],
  Phys.\ Rev.\ D {\bf 97} (2018) 072002
  [arXiv:1710.00971 [hep-ex]].

\bibitem{NA64m}
  S.~N.~Gninenko, N.~V.~Krasnikov and V.~A.~Matveev,
  Phys.\ Rev.\ D {\bf 91} (2015) 095015
  [arXiv:1412.1400 [hep-ph]].

\bibitem{Altmannshofer:2019zhy}
  W.~Altmannshofer, S.~Gori, J.~Mart\'in-Albo, A.~Sousa and M.~Wallbank,
  arXiv:1902.06765 [hep-ph].

\bibitem{Cai:2018nob} 
  C.~Cai, Z.~Kang, H.~H.~Zhang and Y.~P.~Zeng,
  Phys.\ Lett.\ B {\bf 784} (2018) 385.



 \end{thebibliography}
\end{document}